\documentclass[11pt,xcolor=dvipsnames]{article}
\DeclareMathAlphabet{\scr}{U}{rsfs}{m}{n}

\pdfoutput=1
\usepackage{latexsym}
\usepackage{epsfig}
\usepackage[mathscr]{eucal}
\usepackage{amsfonts}
\usepackage{amscd}
\usepackage{cite}
\usepackage{array}
\usepackage{amssymb}
\usepackage{colordvi}
\usepackage[centertags]{amsmath}
\usepackage{enumerate}
\usepackage{graphicx}
\usepackage{booktabs}
\usepackage{theorem}
\usepackage[footnotesize]{caption}
\usepackage{soul}
\usepackage{mcite}
\usepackage{slashed}
\usepackage{xcolor}
\usepackage{bbm}
\usepackage[utf8]{inputenc}
\usepackage{fancyvrb}
\newcommand{\newc}{\newcommand}
\newc{\be}{\begin{equation}}
\newc{\ee}{\end{equation}}
\newc{\bea}{\begin{eqnarray}}
\newc{\eea}{\end{eqnarray}}
\newc{\ol}{\overline}
\newc{\wt}{\widetilde}
\newc{\bs}{\boldsymbol}
\newc{\m}{\mathcal}
\newc{\la}{\langle}
\newc{\ra}{\rangle}

\newcommand{\beq}{\begin{eqnarray}}
\newcommand{\eeq}{\end{eqnarray}}
\newcommand{\bpmatrix}{\begin{pmatrix}}
\newcommand{\epmatrix}{\end{pmatrix}}
\newcommand{\ba}{\begin{array}}
\newcommand{\ea}{\end{array}}

\renewcommand{\ol}{\text{1l}}

\renewcommand{\eqref}[1]{Eq.~(\ref{#1})}

\newcommand{\bc}{\begin{center}}
\newcommand{\ec}{\end{center}}

\newcommand{\gsim}{\raisebox{-0.13cm}{~\shortstack{$>$ \\[-0.07cm]
      $\sim$}}~}
\newcommand{\lsim}{\raisebox{-0.13cm}{~\shortstack{$<$ \\[-0.07cm]
      $\sim$}}~}
\newcommand{\s}{\newline \vspace*{-3.5mm}}

\begin{document}

\title{
\vspace*{-3cm}
\phantom{h} \hfill\mbox{\small KA-TP-39-2016}
\\[1cm]
\textbf{The N2HDM under\\
Theoretical and Experimental Scrutiny\\[4mm]}}

\date{}
\author{
Margarete M\"{u}hlleitner$^{1\,}$\footnote{E-mail:
\texttt{margarete.muehlleitner@kit.edu}} ,
Marco O.~P.~Sampaio$^{2\,}$\footnote{E-mail:
\texttt{msampaio@ua.pt}} ,
Rui Santos$^{3,4\,}$\footnote{E-mail:
  \texttt{rasantos@fc.ul.pt}} ,
Jonas Wittbrodt$^{1,5\,}$\footnote{E-mail:
  \texttt{jonas.wittbrodt@desy.de}}
\\[9mm]
{\small\it
$^1$Institute for Theoretical Physics, Karlsruhe Institute of Technology,} \\
{\small\it 76128 Karlsruhe, Germany}\\[3mm]
{\small\it
$^2$Departamento de F\'{\i}sica, Universidade de Aveiro and CIDMA,} \\
{\small\it Campus de Santiago, 3810-183 Aveiro, Portugal}\\[3mm]
{\small\it
$^3$ISEL -
 Instituto Superior de Engenharia de Lisboa,} \\
{\small \it   Instituto Polit\'ecnico de Lisboa
 1959-007 Lisboa, Portugal} \\[3mm]
{\small\it
$^4$Centro de F\'{\i}sica Te\'{o}rica e Computacional,
    Faculdade de Ci\^{e}ncias,} \\
{\small \it    Universidade de Lisboa, Campo Grande, Edif\'{\i}cio C8
  1749-016 Lisboa, Portugal} \\[3mm]
{\small\it
$^5$Deutsches Elektronen-Synchrotron DESY, Notkestra{\ss}e 85, D-22607
Hamburg, Germany}
}

\maketitle

\begin{abstract}
\noindent
The N2HDM is based on the CP-conserving 2HDM extended by a real scalar
singlet field. Its enlarged parameter space and its fewer
symmetry conditions as compared to supersymmetric models allow for an
interesting phenomenology compatible with current experimental
constraints, while adding to the 2HDM sector
the possibility of Higgs-to-Higgs decays with three different Higgs bosons.
In this paper the N2HDM is subjected to detailed
scrutiny. Regarding the theoretical constraints we implement tests of tree-level
perturbativity and vacuum stability. Moreover, we present, for the
first time, a thorough analysis
of the global minimum of the N2HDM. The model and the theoretical
constraints have been implemented in {\tt ScannerS}, and we provide
{\tt N2HDECAY}, a code based on {\tt HDECAY}, for the computation of
the N2HDM branching ratios and total widths including the
state-of-the-art higher order QCD corrections and off-shell
decays. We then perform an extensive parameter scan in the N2HDM
parameter space, with all theoretical and experimental
constraints applied, and analyse its allowed regions. We find that large
singlet admixtures are still compatible with the Higgs data and investigate
which observables will allow to restrict the singlet nature most
effectively in the next runs of the LHC. Similarly to the 2HDM,
the N2HDM exhibits a wrong-sign
parameter regime, which will be constrained by future Higgs precision
measurements.
\end{abstract}
\thispagestyle{empty}
\vfill
\newpage
\setcounter{page}{1}

\section{Introduction}
The discovery of the Higgs boson by the LHC experiments ATLAS \cite{Aad:2012tfa}
and CMS \cite{Chatrchyan:2012ufa}
not only marked a milestone for elementary particle physics but also
opened the possiblity to search for new physics (NP) in the Higgs
sector itself. Since, so far, a direct discovery of NP in the form of new
particles is missing, the Higgs sector plays an increasingly important role. The
manifestations of NP in the Higgs sector can be manifold
\cite{Englert:2014uua}. An immediate direct signal of NP acting in the
Higgs sector would be the discovery of additional Higgs bosons,
which can be lighter or heavier than the currently known one that has a
mass of 125.09~GeV \cite{Aad:2015zhl}. Indirect signs may appear
through modifications in the Higgs couplings to the Standard Model
(SM) particles and hence through the observables of the 125~GeV Higgs
boson when compared to the SM values. The modifications can be due to
strong dynamics behind electroweak symmetry breaking  (EWSB) like in
composite Higgs models
\cite{Terazawa:1976xx,Terazawa:1979pj,Kaplan:1983fs,Dimopoulos:1981xc,Banks:1984gj,Kaplan:1983sm,Georgi:1984ef,Georgi:1984af,Dugan:1984hq,Giudice:2007fh,Agashe:2004rs,Contino:2006qr}.
In the case of weakly coupled models with extended Higgs sectors, the
SM-like Higgs boson mixes with the other Higgs bosons thus changing
the couplings to the SM particles. Additionally, new non-SM particles,
like {\it e.g.}~the superpartners in supersymmetric extensions, can
contribute to the loop-induced couplings to gluons or
photons. Furthermore, the different particle content and the modified
couplings induce higher order corrections to the Higgs couplings that
can be substantially different from the SM case. Finally, the
additional Higgs bosons open the possibility of Higgs-to-Higgs
decays. These, and possibly invisible decays due to additional lighter
Higgs or other particles that are stable, modify the total width and
hence the branching ratios of the SM-like Higgs boson. \s

With the observed Higgs boson behaving very SM-like
\cite{Khachatryan:2014kca,Aad:2015mxa,Khachatryan:2014jba,Aad:2015gba}
it is clear that any extension of the Higgs sector beyond the SM (BSM)
has to provide at least one CP-even Higgs boson with a mass of 125 GeV
that reproduces the LHC rates. Additional Higgs bosons predicted by
the model have to be compatible with the LHC exclusion
bounds. A strong constraint on NP models is given by the $\rho$
parameter. This singles out models with singlet
or doublet extended Higgs sectors when some
simplicity is required.\footnote{Also models with larger $SU(2)$ multiplets or with
triplets and a custodial $SU(2)$ global symmetry satisfy the $\rho$
parameter constraint, but entail larger and more complex Higgs
sectors.} Doublet extended models are particularly interesting due to
their relation to supersymmetry. In particular, the 2-Higgs-doublet model (2HDM)
\cite{Gunion:1989we,Lee:1973iz,Branco:2011iw} has been
extensively studied and considered as a possible benchmark model in
experimental analyses.  The 2HDM features 5 physical Higgs bosons
that, in the CP-conserving version of the model, are given by 2 CP-even, 1 CP-odd
and 2 charged Higgs bosons. Upon extending the model by a real scalar
singlet field with a ${\mathbb Z}_2$ parity symmetry, there is a
symmetric phase containing a viable Dark
Matter (DM) candidate. This version of the Next-to-Minimal 2HDM
(N2HDM) has been subject to numerous investigations, see {\it
  e.g.}~\cite{He:2008qm,Grzadkowski:2009iz,Logan:2010nw,Boucenna:2011hy,He:2011gc,Bai:2012nv,He:2013suk,Cai:2013zga,Guo:2014bha,Wang:2014elb,Drozd:2014yla,Campbell:2015fra,Drozd:2015gda,vonBuddenbrock:2016rmr},
while in \cite{Chen:2013jvg} the phenomenology of the N2HDM with non-vanishing
vacuum expectation value (VEV) for the singlet field ($\mathbb{Z}_2$
broken phase) has been discussed. In the $\mathbb{Z}_2$ broken phase,
after EWSB the N2HDM Higgs sector
consists of 3 neutral CP-even scalars, 1 CP-odd and 2 charged Higgs
bosons. The Higgs mass eigenstates, which are now superpositions of
the singlet and doublet fields, have an interesting phenomenology that
is not only governed by the mixing properties of the doublet fields
but also by the amount of singlet admixture to the Higgs mass
eigenstates. Thus, the couplings to
SM particles can be diluted to such an extent that light Higgs bosons
are not excluded by Higgs boson searches at LEP, Tevatron and the LHC
in the low-mass range. Such light Higgs bosons then allow for Higgs
decays of the heavier Higgs bosons into a pair of light Higgs states.
Higgs-to-Higgs decays provide alternative production
channels for the heavier Higgs bosons and give access to the trilinear
Higgs self-couplings. Their measurement is crucial for our
understanding of the Higgs mechanism
\cite{Djouadi:1999gv,Djouadi:1999rca,Muhlleitner:2000jj}. Furthermore,
the larger number of parameters, as compared {\it e.g.}~to the 2HDM,
allows for more flexibility in the Higgs sector while being simultaneously in
accordance with the experimental and theoretical constraints. This is
also the case for the Next-to-Minimal Supersymmetric Model
(NMSSM) ~\cite{Fayet:1974pd,Barbieri:1982eh,Dine:1981rt,Nilles:1982dy,Frere:1983ag,Derendinger:1983bz,Ellis:1988er,Drees:1988fc,Ellwanger:1993xa,Ellwanger:1995ru,Ellwanger:1996gw,Elliott:1994ht,King:1995vk,Franke:1995tc,Maniatis:2009re,Ellwanger:2009dp} whose Higgs sector is based on two doublets and one complex
singlet field. Through the SUSY sector the NMSSM encounters even more
parameters.
The NMSSM Higgs potential, however, is
subject to supersymmetric relations.
In particular, the Higgs potential parameters of the two Higgs
doublets in the NMSSM are given in terms of the gauge boson couplings,
so that neither the NMSSM Higgs masses nor the
trilinear Higgs self-couplings can become arbitrarily large. The larger Higgs
spectrum of the NMSSM, with an additional pseudoscalar Higgs, and the
different Higgs self-couplings induce differences in Higgs-to-Higgs
decays as compared to the N2HDM. The supersymmetric
relations furthermore lead to constraints in the Higgs boson
couplings to the SM particles. Therefore differences in the Higgs
rates and also in the coupling patterns, namely in the coupling sum rules, are to be
expected.
The N2HDM, on the other hand, does not have to respect supersymmetry relations
among the masses and couplings. This leads to much more freedom in
the choice of parameters of the model and to very different patterns
in the couplings of the SM-like Higgs boson.
Another class of models that can also provide such extra freedom is
given by the scalar singlet framework where one adds hypyercharge-zero
singlet scalar fields to the SM, {\it i.e.}~without introducing any extra
doublets \cite{Davoudiasl:2004be, vanderBij:2006ne, Datta:1997fx,
  Schabinger:2005ei, BahatTreidel:2006kx, Robens:2015gla,
  Barger:2006sk, Barger:2007im, Barger:2008jx,
  O'Connell:2006wi,Gupta:2011gd, Ahriche:2013vqa,Coimbra:2013qq,
  Chen:2014ask,Profumo:2014opa,Costa:2014qga,Costa:2015llh}. Though in
some of these models~\cite{Costa:2015llh} one can still obtain a rich
phenomenology of Higgs-to-Higgs decays with several Higgs bosons, the
coupling structure of the new Higgs bosons to other SM particles is
typically controlled by only a global suppression factor (relative to
an SM-like Higgs boson). Thus, this provides still less structure than the
N2HDM. Finally, being a model with a 2HDM-like sector, the N2HDM also
contains a richer spectrum with a charged and a CP-odd Higgs boson
that induce different signatures when compared with scalar singlet
models.
The N2HDM
therefore provides an important benchmark model with a very distinct Higgs
boson phenomenology as compared to other commonly studied beyond the
SM extensions.
The potential of various observables to distinguish between
all these models requires a detailed comparison, which is
beyond the scope of this paper and deferred to future work.
\s

The LHC Higgs data constrain possible deviations induced by NP to be
close to the SM case so that only precision measurements allow to
reveal BSM signals in the
Higgs sector. This calls not only for advanced experimental techniques but also
for very precise predictions from the theory side. Thus parameters and
observables have to be computed including higher
order corrections. Moreover the allowed parameter space of the model
has to be evaluated very carefully by checking for consistency with
the relevant theoretical and experimental
constraints. Only for these parameter regions predictions become meaningful and
can be used as guidelines for the experiments.
Constraints from the experimental side arise from the Higgs data. The
N2HDM has to provide at least one SM-like Higgs boson with a mass of
125 GeV. The additional Higgs bosons must comply with the LHC
exclusion limits. Furthermore $B$-physics and low-energy physics
constraints have to be respected as well as the compatibility with the
electroweak precision data. Finally, the symmetric
N2HDM, which features a Dark Matter (DM) candidate has to comply with the
measured value of the relic density\footnote{It is also possible to have a $\mathbb{Z}_2$ symmetric scalar
  singlet model with two CP-even Higgs bosons and a DM
  candidate~\cite{Coimbra:2013qq,Costa:2014qga}. Such a model can be
  made compatible with DM observables and help to stabilise the SM
  potential~\cite{Costa:2014qga}. However, its coupling structure and
  spectrum are, again, simpler than in the N2HDM.}. Theoretical constraints
that have to be fulfilled are: that the Higgs potential is bounded
from below, that the chosen vacuum is a global minimum and that
perturbative unitarity holds. In the N2HDM, the conditions for the
first two requirements can be derived from the literature. There exists, however,
no analysis so far of all the minima of the N2HDM. \s

In this work we will determine the allowed parameter space of the
N2HDM in the broken phase without applying any approximations on the
singlet admixture to the SM-like Higgs boson. Besides taking into
account the experimental constraints, we will, in particular, investigate for the
first time in great detail the conditions on the N2HDM potential that guarantee
tree-level perturbative unitarity, that the vacuum is stable and that the
minimum is the global one. We will present the full analysis of the
global minimum of the N2HDM potential, which was performed for the first time in
\cite{JWittbrodt2016} where more details can be found. We have
implemented the N2HDM in {\tt
  HDECAY}~\cite{Djouadi:1997yw,Butterworth:2010ym}. This code, called
{\tt N2HDECAY}, computes the N2HDM Higgs decay widths and branching
ratios including the state-of-the-art higher order QCD corrections and off-shell
decays. Furthermore, the model has been included in {\tt
  ScannerS}~\cite{Coimbra:2013qq,ScannerS} along with the theoretical
conditions and the available experimental constraints. Then, this
allowed us to perform extensive scans in the parameter space of this
model taking into account the experimental and theoretical
constraints. We will subsequently investigate the features of the
surviving parameter space and the implications for LHC phenomenology. \s

The outline of the paper is as follows. In section \ref{sec:model} we
will introduce the N2HDM together with our notation. Section \ref{sec:constraints}
is dedicated to the description of the theoretical constraints that will be
applied here for the first time in full scrutiny without any
approximations on  the N2HDM Higgs potential. Section
\ref{sec:scans} describes the parameter scan with the applied
constraints. Section \ref{sec:pheno} is dedicated to the
phenomenological analysis. Our conclusions are collected in section~\ref{sec:concl}. \s

\section{The N2HDM Higgs Sector \label{sec:model}}
\setcounter{equation}{0}
The N2HDM is based on the CP-conserving (or real) 2HDM with a
softly broken
$\mathbb{Z}_2$ symmetry extended by a real singlet field $\Phi_S$.
The extension
of the 2HDM by real scalar singlet that does not acquire a VEV
provides a viable DM candidate
\cite{He:2008qm,Grzadkowski:2009iz,Logan:2010nw,Boucenna:2011hy,He:2011gc,Bai:2012nv,He:2013suk,Cai:2013zga,Wang:2014elb,Drozd:2014yla,Campbell:2015fra,vonBuddenbrock:2016rmr}. In
\cite{Chen:2013jvg} the phenomenology of the N2HDM with non-vanishing
VEV for the singlet field has been discussed by applying some approximations. In
particular, the possibility of a singlet admixture to the 125 GeV
Higgs boson has been neglected. In the following no such assumptions will
be imposed on the N2HDM potential. In terms of the two $SU(2)_L$ Higgs
doublets $\Phi_1$ and $\Phi_2$ and the singlet field
$\Phi_S$, the N2HDM potential is given by
\beq
V &=& m_{11}^2 |\Phi_1|^2 + m_{22}^2 |\Phi_2|^2 - m_{12}^2 (\Phi_1^\dagger
\Phi_2 + h.c.) + \frac{\lambda_1}{2} (\Phi_1^\dagger \Phi_1)^2 +
\frac{\lambda_2}{2} (\Phi_2^\dagger \Phi_2)^2 \nonumber \\
&& + \lambda_3
(\Phi_1^\dagger \Phi_1) (\Phi_2^\dagger \Phi_2) + \lambda_4
(\Phi_1^\dagger \Phi_2) (\Phi_2^\dagger \Phi_1) + \frac{\lambda_5}{2}
[(\Phi_1^\dagger \Phi_2)^2 + h.c.] \nonumber \\
&& + \frac{1}{2} m_S^2 \Phi_S^2 + \frac{\lambda_6}{8} \Phi_S^4 +
\frac{\lambda_7}{2} (\Phi_1^\dagger \Phi_1) \Phi_S^2 +
\frac{\lambda_8}{2} (\Phi_2^\dagger \Phi_2) \Phi_S^2 \;.
\label{eq:n2hdmpot}
\eeq
The first two lines describe the 2HDM part of the N2HDM while the last line contains the contribution of the singlet field $\Phi_S$.
This potential is obtained by imposing two $\mathbb{Z}_2$ symmetries
on the scalar potential. The first one, called $\mathbb{Z}_2$,
\begin{align}
  \Phi_1 \to \Phi_1\;, \quad \Phi_2 \to - \Phi_2\;, \quad \Phi_S \to \Phi_S \label{eq:2HDMZ2}
\end{align}
is the trivial generalisation of the usual 2HDM $\mathbb{Z}_2$
symmetry to the N2HDM.
It is softly broken by the term involving $m_{12}^2$ and can be extended to the Yukawa sector to guarantee the absence of tree-level Flavour Changing Neutral Currents (FCNC).
The second $\mathbb{Z}_2$ symmetry, $\mathbb{Z}'_2$, is
\begin{align}
  \Phi_1 \to \Phi_1\;, \quad \Phi_2 \to \Phi_2\;, \quad \Phi_S \to -\Phi_S \label{eq:singZ2}
\end{align}
and is not explicitly broken.
If $\Phi_S$ does not acquire a VEV this second $\mathbb{Z}_2$ symmetry
will give rise to a conserved ``darkness'' quantum number and to the
appearance of a dark matter candidate. If $\Phi_S$ acquires a VEV this
quantum number is no longer conserved and there is mixing among all
CP-even neutral particles. This same behaviour is still possible if
$m_{12}^2=0$, where both $\mathbb{Z}_2$ symmetries in the potential
are exact but need to be spontaneously broken. We will not consider
such model further in this study. The two $\mathbb{Z}_2$ quantum
numbers assigned to the scalars in the model are shown in table~\ref{discrete}.

\begin{table}[h!]
\begin{center}
  \begin{tabular}{l|ccc}
  \toprule
   & $\Phi_1$ & $\Phi_2$ & $\Phi_S$  \\
   \midrule
$\mathbb{Z}_2$ ({\rm explicitly\hspace{1mm}broken, softly}) & $+$ & $-$ & $+$ \\
$\mathbb{Z}'_2$ ({\rm spontaneously broken}) &  $+$ & $+$ & $-$  \\
\bottomrule
   \end{tabular}
\end{center}
\vspace*{-0.4cm}
  \caption{$\mathbb{Z}_2$ and $\mathbb{Z}'_2$ assignments for
    the scalars fields in the model.
 }
  \label{discrete}
 \end{table}


After EWSB the two doublet fields
acquire the real VEVs $v_1$ and $v_2$ and the singlet field a real VEV
$v_S$. They can be parametrised as
\beq
\Phi_1 = \left( \begin{array}{c} \phi_1^+ \\ \frac{1}{\sqrt{2}} (v_1 +
    \rho_1 + i \eta_1) \end{array} \right) \;, \quad
\Phi_2 = \left( \begin{array}{c} \phi_2^+ \\ \frac{1}{\sqrt{2}} (v_2 +
    \rho_2 + i \eta_2) \end{array} \right) \;, \quad
\Phi_S = v_S + \rho_S \;, \label{eq:n2hdmfields}
\eeq
in terms of the charged complex fields $\phi_i^+$ ($i=1,2$) and the real neutral
CP-even and CP-odd fields $\rho_{I}$ ($I=1,2,S$) and $\eta_i$, respectively.
Requiring the potential to be minimized at the VEV leads to three
minimum conditions given by
\beq
\frac{v_2}{v_1} m_{12}^2 - m_{11}^2 &=& \frac{1}{2} (v_1^2 \lambda_1 +
v_2^2 \lambda_{345} + v_S^2 \lambda_7) \label{eq:n2hdmmin1} \\
\frac{v_1}{v_2} m_{12}^2 - m_{22}^2 &=& \frac{1}{2} (v_1^2 \lambda_{345} +
v_2^2 \lambda_2 + v_S^2 \lambda_8) \label{eq:n2hdmmin2} \\
- m_S^2 &=& \frac{1}{2} (v_1^2 \lambda_7 + v_2^2 \lambda_8 + v_S^2
\lambda_6) \;, \label{eq:n2hdmmin3}
\eeq
with
\beq
\lambda_{345} \equiv \lambda_3 + \lambda_4 + \lambda_5 \;.
\label{eq:l345}
\eeq
Replacing the doublet and singlet fields in the Higgs potential by
the parametrisations (\ref{eq:n2hdmfields}) the mass matrices in the
gauge basis are obtained from the second derivatives with respect to
the fields in the gauge basis. Due to charge and CP conservation the $7\times 7$
mass matrix decomposes into three blocks: the $2\times 2$ matrix for
the charged Higgs bosons, the $2\times 2$ matrix for the CP-odd fields
and the $3\times 3$ matrix for the CP-even states.
Introducing a real singlet field with a VEV, the charged and
pseudoscalar sectors of the model remain unchanged with respect to the
2HDM. Consequently, as in the 2HDM, the charged and pseudoscalar mass
matrices can be diagonalised by the rotation matrix
\beq
R_\beta = \left( \begin{array}{cc} c_\beta & s_\beta \\ - s_\beta &
    c_\beta \end{array} \right) \;,
\eeq
with $t_\beta$ defined as
\beq
t_\beta = \frac{v_2}{v_1}
\eeq
and
\beq
v^2 = v_1^2 + v_2^2 \;.
\eeq
We have introduced the SM VEV $v \approx 246$~GeV and the
abbreviations $\sin x \equiv s_x$, $\cos x \equiv c_x$ and $\tan x \equiv t_x$.
This yields the massless charged and neutral would-be Goldstone
bosons $G^\pm$ and $G^0$, the charged Higgs mass
eigenstates $H^\pm$ and the pseudoscalar mass eigenstate $A$. \s

Due to the additional real singlet field, the CP-even neutral
sector of the N2HDM is changed with respect to the 2HDM. Instead of a
$2\times 2$ mass matrix we now have a $3\times 3$ matrix. In
the basis $(\rho_1, \rho_2, \rho_S)$ it can be cast into the form
\beq
M_{\text{scalar}}^2 = \left( \begin{array}{ccc} \lambda_1 c_\beta^2 v^2 + t_\beta
    m_{12}^2 & \lambda_{345} c_\beta s_\beta v^2 - m_{12}^2 &
    \lambda_7 c_\beta v v_S \\ \lambda_{345} c_\beta s_\beta v^2 - m_{12}^2 &
    \lambda_2 s_\beta^2 v^2 + m_{12}^2/t_\beta & \lambda_8 s_\beta v
    v_S \\ \lambda_7 c_\beta v v_S & \lambda_8 s_\beta v v_S &
    \lambda_6 v_S^2 \end{array} \right) \;.
\label{eq:neutralmassmatrix}
\eeq
In Eq.~(\ref{eq:neutralmassmatrix}) we have used
Eqs.~(\ref{eq:n2hdmmin1})-(\ref{eq:n2hdmmin3}),  to
trade the mass parameters $m_{11}^2$, $m_{22}^2$ and $m_S^2$ for $v$,
$t_\beta$ and $v_S$. The mass matrix can be
diagonalised by an orthogonal matrix $R$ which we parametrise as
\beq
R =\left( \begin{array}{ccc}
c_{\alpha_1} c_{\alpha_2} & s_{\alpha_1} c_{\alpha_2} & s_{\alpha_2}\\
-(c_{\alpha_1} s_{\alpha_2} s_{\alpha_3} + s_{\alpha_1} c_{\alpha_3})
& c_{\alpha_1} c_{\alpha_3} - s_{\alpha_1} s_{\alpha_2} s_{\alpha_3}
& c_{\alpha_2} s_{\alpha_3} \\
- c_{\alpha_1} s_{\alpha_2} c_{\alpha_3} + s_{\alpha_1} s_{\alpha_3} &
-(c_{\alpha_1} s_{\alpha_3} + s_{\alpha_1} s_{\alpha_2} c_{\alpha_3})
& c_{\alpha_2}  c_{\alpha_3}
\end{array} \right)
\label{eq:mixingmatrix}
\eeq
in terms of the mixing angles $\alpha_1$ to $\alpha_3$. Without loss
of generality the angles can be chosen in the range
\beq
- \frac{\pi}{2} \le \alpha_{1,2,3} < \frac{\pi}{2} \;.
\eeq
The matrix $R$ rotates the interaction basis $(\rho_1, \rho_2, \rho_S)$ into the physical mass eigenstates $H_1$, $H_2$
and $H_3$,
\beq
\left( \begin{array}{c} H_1 \\ H_2 \\ H_3 \end{array} \right) = R
\left( \begin{array}{c} \rho_1 \\ \rho_2 \\ \rho_S \end{array} \right)
\eeq
and diagonalises the mass matrix $M_{\text{scalar}}^2$,
\beq
R M_{\text{scalar}}^2 R^T = \mbox{diag}(m_{H_1}^2,m_{H_2}^2,m_{H_3}^2) \;.
\eeq
We use a convention where the mass eigenstates are ordered by ascending mass as
\beq
m_{H_1} < m_{H_2} < m_{H_3} \;.
\eeq
In total, the N2HDM  is described by 12 independent real parameters.  We choose
as many parameters with physical meaning as possible. We use the minimisation
conditions to trade $m_{11}^2$, $m_{22}^2$ and $m_S^2$ for the SM VEV $v$,
$t_\beta$ and $v_S$ and replace the quartic couplings by the
physical masses and the mixing angles. The soft $\mathbb{Z}_2$ breaking parameter $m_{12}^2$ is kept as an independent parameter.
Thus, we use the following set of input parameters
\beq
\alpha_1 \; , \quad \alpha_2 \; , \quad \alpha_3 \; , \quad t_\beta \;, \quad v \; ,
\quad v_S \; , \quad m_{H_{1,2,3}} \;, \quad m_A \;, \quad m_{H^\pm}
\;, \quad m_{12}^2 \;. \label{eq:n2hdminputpars}
\eeq
In appendix~\ref{app:lamrels} we provide expressions for the quartic
couplings in terms of these input parameters. \s
\begin{table}[b!]
\begin{center}
 \begin{tabular}{cc}
\toprule
\multicolumn{2}{c}{$c(H_i VV)$} \\
\midrule
$H_1$ & $c_{\alpha_2} c_{\beta-\alpha_1}$ \\
$H_2$ & $-c_{\beta-\alpha_1} s_{\alpha_2} s_{\alpha_3} + c_{\alpha_3}
s_{\beta-\alpha_1}$ \\
$H_3$ & $-c_{\alpha_3} c_{\beta-\alpha_1} s_{\alpha_2} - s_{\alpha_3}
s_{\beta-\alpha_1}$ \\
\bottomrule
\end{tabular}
 \caption{The effective couplings of the neutral CP-even N2HDM Higgs bosons
   $H_i$ to the massive gauge bosons $V=W,Z$. \label{tab:gaugecoupn2hdm}}
\end{center}
\end{table}

The singlet field $\rho_S$ does not directly couple to the SM
particles. Therefore, any change in the tree-level Higgs couplings compared to the
2HDM is due to the mixing of the three neutral fields
$\rho_I$. Any coupling not involving the CP-even neutral
Higgs bosons remains unchanged compared to the 2HDM and can be found {\it
  e.g.}~in \cite{Branco:2011iw}. We now provide the
couplings of the N2HDM Higgs bosons $H_i$ relevant for Higgs
decays. We introduce the Feynman rules for the Higgs couplings $H_i$ to the
massive gauge bosons $V\equiv W,Z$ as
\beq
i \, g_{\mu\nu} \, c(H_i VV) \, g_{H^{\text SM} VV} \;, \label{eq:gaugecoupdef}
\eeq
where $g_{H^{\text SM} VV}$ denotes the SM Higgs coupling factor. In terms of
the gauge boson masses $M_W$ and $M_Z$, the $SU(2)_L$ gauge coupling
$g$ and the Weinberg angle
$\theta_W$ it is given by
\beq
g_{H^{\text SM} VV} = \left\{ \begin{array}{ll} g M_W & \quad
    \mbox{for } V=W \\
 g M_Z/\cos\theta_W & \quad \mbox{for } V=Z \end{array} \right. \;.
\eeq
We obtain
\beq
c(H_i VV) = c_\beta R_{i1} + s_\beta R_{i2} \label{eq:n2hdmgaugecoup}
\eeq
for the effective couplings defined by Eq.~(\ref{eq:gaugecoupdef}).
Replacing the $R_{ij}$ by their
parametrisation in terms of the mixing angles yields the effective
couplings in table~\ref{tab:gaugecoupn2hdm}. \s

In order to avoid tree-level FCNCs we extend the
$\mathbb{Z}_2$ symmetry (\ref{eq:2HDMZ2}) to the Yukawa sector. This
leads to the same four types of doublet couplings to the
fermions  as in the 2HDM. We show these types in table \ref{tab:types}.
Consequently, the CP-even
$H_i$ Yukawa couplings take the same form as the
Yukawa couplings of the 2HDM. With the N2HDM Yukawa Lagrangian
\beq
{\cal L}_Y = - \sum_{i=1}^3 \frac{m_f}{v} c(H_i ff) \,
\bar{\psi}_f \psi_f H_i
\label{eq:lyukn2hdm}
\eeq
we obtain the effective coupling factors $c(H_i ff)$ in
terms of the mixing matrix elements $R_{ij}$ and the mixing angle $\beta$
provided in table~\ref{tab:yukcoup}.
Replacing the $R_{ij}$ by their parametrisation in terms
of the $\alpha_i$ this results in the effective coupling expressions given for
type I and II in table \ref{tab:effyukn2hdm}. \s

\begin{table}[t!]
\begin{center}
\begin{tabular}{rccc| ccccc} \toprule
  & $u$-type & $d$-type & leptons & $Q$ & $u_R$ & $d_R$ & $L$ & $l_R$ \\
  \midrule
type I & $\Phi_2$ & $\Phi_2$ & $\Phi_2$ & $+$ & $-$ & $-$ & $+$ & $-$  \\
type II & $\Phi_2$ & $\Phi_1$ & $\Phi_1$ & $+$ & $-$ & $+$ & $+$ & $-$ \\
lepton-specific & $\Phi_2$ & $\Phi_2$ & $\Phi_1$ & $+$ & $-$ & $+$ & $+$ & $-$\\
flipped & $\Phi_2$ & $\Phi_1$ & $\Phi_2$ & $+$ & $-$ & $-$ & $+$ & $+$\\ \bottomrule
\end{tabular}
\caption{In the three leftmost columns, the four Yukawa types of the $\mathbb{Z}_2$-symmetric 2HDM
  are defined by the Higgs doublet that couples to each kind of fermion. The five rightmost columns show the corresponding $\mathbb{Z}_2$ parity assignments for the fermions. $Q$ and $L$ are the quark and lepton doublets, respectively, $u_R$ is the up-type quark singlet, $d_R$ is the down-type quark singlet and $l_R$ is the lepton singlet. Observe that, in addition, all gauge bosons are $\mathbb{Z}_2$-even for all types. The $\mathbb{Z}'_2$ parities are not included in the table because all fields, except for $\Phi_S$, are even. \label{tab:types}}
\end{center}
\vspace*{-1cm}
\end{table}
The Feynman rule for the $H_i$ coupling to the pseudoscalar $A$ and the
$Z$ boson is given by
\beq
\lambda_\mu (H_i ZA) = \frac{\sqrt{g^2 + g'^2}}{2} (p_{H_i} - p_A)_\mu \,
\tilde{c}(H_i V H) \;, \label{eq:couphiaz}
\eeq
where $g'$ denotes the $U(1)_Y$ gauge coupling and $p_A$ and $p_{H_i}$,
the four-momenta of the pseudoscalar and the $H_i$, are both taken as
incoming. The tilde over the coupling factor indicates that it is not
an effective coupling in the sense that it is not normalized to a
corresponding SM coupling, since there is no SM counterpart. The
corresponding Feynman rule for the $H_i$ coupling to the
charged pair $H^\pm$ and $W^\mp$ reads
\beq
\lambda_\mu (H_i W^\mp H^\pm) = \mp \frac{g}{2} (p_{H_i} -
p_{H^\pm})_\mu \, \tilde{c}(H_i V H) \;, \label{eq:couphihw}
\eeq
where $p_{H^\pm}$ denotes the four-momentum of $H^\pm$ and again all
momenta are taken as incoming. The coupling
factors $\tilde{c} (H_i VH)$ are provided in table~\ref{tab:couphivh}. \s

The trilinear Higgs self-couplings relevant for the Higgs decays into a
pair of lighter Higgs bosons are quite lengthy and deferred to
appendix~\ref{app:trilcoup}. We have used these Feynman rules to
implement the N2HDM in {\tt HDECAY v6.51} \cite{}. The resulting code
{\tt N2HDECAY}
calculates all N2HDM Higgs boson decay widths and branching ratios
including state-of-the-art higher order QCD corrections and
off-shell decays. Electroweak corrections, which in contrast to the QCD
corrections cannot be taken over from the SM, have been consistently
neglected. The program can be downloaded from the url:\\
\centerline{https://itp.kit.edu/$\sim$maggie/N2HDECAY}
\begin{table}[b!]
\begin{center}
  \begin{tabular}{rccc} \toprule
& $u$-type & $d$-type & leptons \\ \midrule
type I & $\frac{R_{i2}}{s_\beta}$
& $\frac{R_{i2}}{s_\beta}$ &
$\frac{R_{i2}}{s_\beta}$ \\
type II & $\frac{R_{i2}}{s_\beta} $
& $\frac{R_{i1}}{c_\beta} $ &
$\frac{R_{i1}}{c_\beta} $ \\
lepton-specific & $\frac{R_{i2}}{s_\beta}$
& $\frac{R_{i2}}{s_\beta}$ &
$\frac{R_{i1}}{c_\beta}$ \\
flipped & $\frac{R_{i2}}{s_\beta}$
& $\frac{R_{i1}}{c_\beta}$ &
$\frac{R_{i2}}{s_\beta}$ \\ \bottomrule
\end{tabular}
\caption{Coupling coefficients $c(H_i ff)$ of the Yukawa couplings of
   the N2HDM Higgs bosons $H_i$ as defined in
   Eq.~(\ref{eq:lyukn2hdm}). \label{tab:yukcoup}}
\end{center}
\end{table}
\begin{table}
\begin{center}
  \begin{tabular}{cccc}
\multicolumn{4}{c}{Type I} \\ \toprule
$c(H_i ff)$ & $u$ & $d$ & $l$ \\ \midrule
$H_1$ & $(c_{\alpha_2} s_{\alpha_1} )/s_\beta$ & $(c_{\alpha_2}
s_{\alpha_1}) / s_\beta$ & $(c_{\alpha_2} s_{\alpha_1})/s_\beta$ \\
$H_2$ & $(c_{\alpha_1} c_{\alpha_3} - s_{\alpha_1} s_{\alpha_2}
s_{\alpha_3})/s_\beta$ & $(c_{\alpha_1} c_{\alpha_3}- s_{\alpha_1}
s_{\alpha_2} s_{\alpha_3})/s_\beta$ & $(c_{\alpha_1} c_{\alpha_3}-
s_{\alpha_1} s_{\alpha_2} s_{\alpha_3})/s_\beta$ \\
$H_3$ & $-(c_{\alpha_1} s_{\alpha_3} + c_{\alpha_3} s_{\alpha_1}
s_{\alpha_2} )/s_\beta$ & $-(c_{\alpha_1} s_{\alpha_3} + c_{\alpha_3}
s_{\alpha_1} s_{\alpha_2} )/s_\beta$ & $-(c_{\alpha_1} s_{\alpha_3} +
c_{\alpha_3} s_{\alpha_1} s_{\alpha_2}) /s_\beta$ \\ \bottomrule\\
\multicolumn{4}{c}{Type II} \\ \toprule
$c(H_i ff)$ & $u$ & $d$ & $l$ \\ \midrule
$H_1$ & $(c_{\alpha_2} s_{\alpha_1} )/s_\beta$ & $(c_{\alpha_1}
c_{\alpha_2}) / c_\beta$ & $(c_{\alpha_1} c_{\alpha_2})/c_\beta$ \\
$H_2$ & $(c_{\alpha_1} c_{\alpha_3} - s_{\alpha_1} s_{\alpha_2}
s_{\alpha_3})/s_\beta$ & $-(c_{\alpha_3} s_{\alpha_1}+ c_{\alpha_1}
s_{\alpha_2} s_{\alpha_3})/c_\beta$ & $-(c_{\alpha_3} s_{\alpha_1}+
c_{\alpha_1} s_{\alpha_2} s_{\alpha_3})/c_\beta$ \\
$H_3$ & $-(c_{\alpha_1} s_{\alpha_3} + c_{\alpha_3} s_{\alpha_1}
s_{\alpha_2} )/s_\beta$ & $(s_{\alpha_1} s_{\alpha_3} - c_{\alpha_1}
c_{\alpha_3} s_{\alpha_2} )/c_\beta$ & $(s_{\alpha_1} s_{\alpha_3} -
c_{\alpha_1} c_{\alpha_3} s_{\alpha_2}) /c_\beta$ \\ \bottomrule
\end{tabular}
\caption{The effective Yukawa couplings $c(H_i ff)$ of the N2HDM Higgs
   bosons $H_i$, as defined in Eq.~(\ref{eq:lyukn2hdm}), in type I and
   type II. \label{tab:effyukn2hdm}}
\end{center}
\end{table}
\begin{table}[h!]
\begin{center}
\begin{tabular}{cc}
\toprule
\multicolumn{2}{c}{$\tilde{c}(H_i VH)$} \\ \midrule
$H_1$ & $- c_{\alpha_2} s_{\beta-\alpha_1}$ \\
$H_2$ & $s_{\beta-\alpha_1} s_{\alpha_2} s_{\alpha_3} + c_{\alpha_3}
c_{\beta-\alpha_1}$ \\
$H_3$ & $c_{\alpha_3} s_{\beta-\alpha_1} s_{\alpha_2} - s_{\alpha_3}
c_{\beta-\alpha_1}$ \\ \bottomrule
\end{tabular}
\caption{The coupling factors $\tilde{c} (H_i VH)$ as defined in the
 Feynman rules Eqs.~(\ref{eq:couphiaz}) and (\ref{eq:couphihw})
 for the $H_i$ couplings to a pair of Higgs and gauge bosons. \label{tab:couphivh}}
\end{center}
\end{table}
In appendix \ref{app:n2hdecay} a short description of the
program with a sample input and output file is provided. \s

We finally note that by letting
$\alpha_{2,3} \to 0$ and
$\alpha_1 \to \alpha + \pi/2$ the N2HDM approaches the limit of a 2HDM with an added decoupled singlet. In the 2DHM, the
mixing angle $\alpha$ diagonalises the $2\times2$ mass matrix in the CP-even
Higgs sector leading to the two CP-even mass eigenstates $h$ and $H$,
respectively. The shift by $\pi/2$ in the limit is necessary to match the usual 2HDM convention. Hence,
\beq
\mbox{N2HDM } \rightarrow \mbox{ 2HDM } \; \Longleftrightarrow \;
\left\{ \begin{array}{lcl}
\alpha_1 &\to& \alpha +\frac{\pi}{2} \\
\alpha_2 &\to& 0 \\
\alpha_3 &\to& 0
\end{array} \right. \;. \label{eq:2hdmlimit}
\eeq

\section{Theoretical Constraints \label{sec:constraints}}
In this section we investigate the conditions on the N2HDM imposed by
theoretical considerations. These are the requirements for tree-level perturbative
unitarity, that the vacuum is stable and that it is the
global minimum of the scalar potential. For the first two requirements the corresponding conditions in the
N2HDM can be derived from the literature. In the following we will summarize them before
presenting the analysis of the stationary points of
the N2HDM ({\it cf.}~also \cite{JWittbrodt2016}). The model, along with the
theoretical conditions, has been implemented in {\tt ScannerS}. This
allows us to perform extensive scans in the parameter space of the
N2HDM taking into account both the experimental and the theoretical
constraints, as will be described in detail in section~\ref{sec:scans}.

\subsection{Tree-level Perturbative Unitarity  \label{sec:perturb}}
Tree-level perturbative unitarity is ensured by requiring that the
eigenvalues of the $2\to 2$ scalar scattering matrix are below an
absolute upper value given by $8\pi$ \cite{Horejsi:2005da}. It can be useful to
impose a limit smaller than $8\pi$ at tree level to safeguard for possible
large enhancements of the scalar couplings through higher order
corrections. Following the procedure and notation of
\cite{Horejsi:2005da}, we have calculated the full $2\to 2$ scattering
matrix of the fields in the gauge basis,
\beq
H_1^\pm, H_2^\pm, \rho_1, \rho_2, \rho_S, \eta_1 \; \mbox{and} \;
\eta_2 \;.
\eeq
The resulting matrix is block diagonal. The
block matrices not containing $\rho_S$ have the eigenvalues
\beq
b_\pm &=& \frac{1}{2} \left( \lambda_1 + \lambda_2 \pm
 \sqrt{(\lambda_1-\lambda_2)^2 + 4 \lambda_5^2} \right)\\
c_\pm &=& \frac{1}{2} \left( \lambda_1 + \lambda_2 \pm
 \sqrt{(\lambda_1-\lambda_2)^2 + 4 \lambda_4^2} \right)\\
e_1 &=& \lambda_3 + 4 \lambda_4 - 3 \lambda_5 \\
e_2 &=& \lambda_3 - \lambda_5 \\
f_+ &=& \lambda_3 + 2\lambda_4 + 3 \lambda_5 \\
f_- &=& \lambda_3 + \lambda_5 \\
f_1 &=& \lambda_3 + \lambda_4 \\
p_1 &=& \lambda_3 - \lambda_4 \;.
\eeq
These are the same eigenvalues as found in the 2HDM. The new
contributions due to the singlet field yield the eigenvalues
\beq
s_1 &=& \lambda_7 \\
s_2 &=& \lambda_8
\eeq
and the eigenvalues $1/2\, a_{1,2,3}$, where $a_{1,2,3}$ are the real roots of the
cubic polynomial
\beq
&&4\left(-27 \lambda_1 \lambda_2 \lambda_6 + 12 \lambda_3^2 \lambda_6 + 12
\lambda_3 \lambda_4 \lambda_6 + 3 \lambda_4^2 \lambda_6 + 6 \lambda_2
\lambda_7^2 - 8 \lambda_3 \lambda_7 \lambda_8 - 4 \lambda_4 \lambda_7
\lambda_8 + 6 \lambda_1 \lambda_8^2 \right) \nonumber\\
&&+ x \left(36 \lambda_1 \lambda_2 -
16\lambda_3^2 - 16\lambda_3 \lambda_4 - 4 \lambda_4^2 + 18 \lambda_1
\lambda_6 + 18 \lambda_2 \lambda_6 - 4\lambda_7^2 - 4\lambda_8^2\right)
\nonumber\\
&& +x^2
\left(-6 (\lambda_1 + \lambda_2) -3 \lambda_6\right) + x^3 \;. \label{eq:polynomial}
\eeq
Not all of the above eigenvalues are independent. As noted in
\cite{Horejsi:2005da}, we have
\beq
3 f_1 &=& p_1 + e_1 + f_+ \\
3 e_2 &=& 2 p_1 + e_1 \\
3 f_- &=& 2p_1 + f_+ \;.
\eeq
The conditions on $f_1$, $e_2$ and $f_-$ can therefore be dropped,
as they follow from the conditions on $p_1$, $e_1$ and $f_+$. Since
$\lambda_1, \lambda_2 >0$ is necessary for the potential to be bounded
from below (see Eqs.~(\ref{eq:omega1}) and (\ref{eq:omega2})) we
obtain
\beq
|c_+| &>& |c_-| \\
|b_+| &>& |b_-| \;.
\eeq
The resulting conditions for tree-level perturbative unitarity are thus
given by
\beq
|\lambda_3 - \lambda_4| &<& 8 \pi \label{eq:ev1}\\
|\lambda_3 + 2 \lambda_4 \pm 3 \lambda_5| &<& 8 \pi \\
\left| \frac{1}{2} \left( \lambda_1 + \lambda_2 + \sqrt{(\lambda_1 -
     \lambda_2)^2 + 4 \lambda_4^2}\right) \right| &<& 8\pi \\
\left| \frac{1}{2} \left( \lambda_1 + \lambda_2 + \sqrt{(\lambda_1 -
     \lambda_2)^2 + 4 \lambda_5^2}\right) \right| &<& 8\pi \\
|\lambda_7|\,,\;|\lambda_8|&<&8\pi\\
\frac{1}{2}|a_{1,2,3}| &<& 8\pi \;, \label{eq:ev5}
\eeq
where $a_{1,2,3}$ are the real roots of Eq.~(\ref{eq:polynomial}).


\subsection{Boundedness from below \label{sec:bounded}}
We consider
the potential to be bounded from below in the strong sense, which
means that the potential is required to be strictly positive as the
fields approach infinity. The corresponding necessary and sufficient
conditions have been given in \cite{Klimenko:1984qx} and translated to
our notation in \cite{JWittbrodt2016}. They depend on the discriminant
\beq
D = \text{min}(\lambda_4-|\lambda_5|\,,\;0)\;.
\eeq
The allowed region is given by
\beq
\Omega_1 \cup \Omega_2
\eeq
with
\beq
\Omega_1 &=& \Bigg\{ \lambda_1, \lambda_2, \lambda_6 > 0; \sqrt{\lambda_1 \lambda_6} +
\lambda_7 > 0; \sqrt{\lambda_2 \lambda_6} + \lambda_8 > 0; \nonumber
\\
&& \sqrt{\lambda_1 \lambda_2} + \lambda_3 + D > 0; \lambda_7 +
\sqrt{\frac{\lambda_1}{\lambda_2}} \lambda_8 \ge 0 \Bigg\} \label{eq:omega1}
\eeq
and
\beq
\Omega_2 &=& \Bigg\{ \lambda_1, \lambda_2, \lambda_6 > 0; \sqrt{\lambda_2
  \lambda_6} \ge
\lambda_8 > -\sqrt{\lambda_2 \lambda_6}; \sqrt{\lambda_1 \lambda_6} > -
\lambda_7 \ge \sqrt{\frac{\lambda_1}{\lambda_2}} \lambda_8; \nonumber
\\
&&
\sqrt{(\lambda_7^2 - \lambda_1 \lambda_6)(\lambda_8^2 -\lambda_2
  \lambda_6)} > \lambda_7 \lambda_8 - (D+\lambda_3) \lambda_6 \Bigg\}
\;.
\label{eq:omega2}
\eeq

\subsection{Global Minimum Conditions for the N2HDM
  Potential \label{sec:global}}
In general, the minimum needs not be the global minimum, if the tunnelling
time for the vacuum to tunnel into the global minimum
\cite{Coleman:1977py,Callan:1977pt} is larger than
the age of universe. As the calculation of the tunnelling time is
beyond the scope of this work, we do not discuss such metastable vacua
and restrict to the stronger requirement that the vacuum is at the global minimum. While for the 2HDM it has been proven \cite{Ferreira:2004yd}
that the existence of a
normal minimum of the form Eq.~(\ref{eq:n2hdmfields}) precludes the
existence of a deeper charge- or CP-breaking minimum, this does not
generalise to the N2HDM\@. Counter-examples that underline this statement
can be found in the appendix of \cite{JWittbrodt2016}. For the
analysis of the global N2HDM minimum we therefore have to include the
possibility of CP- and charge-breaking minima. We consider
the most general constant field configuration, where all fields are real,
\beq
\langle \Phi_1 \rangle = \left( \begin{array}{c} 0 \\ v_1 \end{array}
\right) \;, \quad
\langle \Phi_2 \rangle = \left( \begin{array}{c} v_{\text{cb}} \\ v_2
    + i v_{\text{cp}} \end{array} \right) \;, \quad
\langle \Phi_S \rangle = v_S \;. \label{eq:fieldconfig}
\eeq
Here we have already exploited the $SU(2)_L \times U(1)_Y$ gauge
symmetry to eliminate four degrees of freedom. Any other possible
constant field configuration of the N2HDM
can be projected onto this one through a gauge transformation. By
$v_{\text{cb}}$ we denote the charge-breaking and by $v_{\text{cp}}$
the CP-breaking constant fields. In the following we will refer to the
constant fields as VEVs although this is technically only correct if the
configuration describes a minimum of the scalar potential.
By expanding the field configuration Eq.~(\ref{eq:fieldconfig}) in the
potential one observes a set of $\mathbb{Z}_2$ symmetries for the real
fields we are using, and we can thus choose, without loss of
generality, all VEVs except for $v_2$ to be positive. We then proceed
to find all possible stationary points of the N2HDM. This detailed
analysis is presented in appendix~\ref{sec:appglobal}. \s

We want the minimum to conserve both electric charge and CP and to give rise to
three CP-even massive scalars. From the vacuum structure point of view
this means that $v_1 \neq 0$, $v_2 \neq 0$ and $v_S \neq 0$
while $v_{\text{cb}}=0$ and $v_{\text{cp}}=0$ at the chosen minimum. In order to ensure
that this is the global minimum we proceed as follows:
\begin{itemize}
\item
We choose the model parameters such that there is a minimum with $v_1 \neq 0$, $v_2 \neq 0$, $v_S \neq 0$, $v_{\text{cb}}=0$, $v_{\text{cp}}=0$
and $246 \, \text{GeV} = v = \sqrt{v_1^2 + v_2^2}$.
\item
Using the stationarity conditions presented in appendix~\ref{sec:appglobal}
we look for all other possible stationary points of the potential with
this set of parameters.
\item
We discard all sets of parameters for which we find a stationary point below
the minimum.
\end{itemize}
This procedure leads to a global minimum that is CP-conserving, preserves
electric charge and allows for the singlet to mix with the CP-even scalars
from the doublets.
We note here that, in the N2HDM, the classification of the possible types of minima, in terms of preservation or breaking of electric charge or CP, follows that of the 2HDM regardless of the value of $v_S$ -- see also appendix~\ref{sec:appglobal}.

\section{The Parameter Scan \label{sec:scans}}
In order to perform phenomenological analyses we need viable parameter
points, {\it
  i.e.}~points in agreement with theoretical and experimental
constraints. To obtain these points we use the program {\tt ScannerS}
to perform extensive scans in the N2HDM parameter space and check for
compatibility with the constraints. We denote the
discovered SM-like Higgs boson with a mass of \cite{Aad:2015zhl}
\beq
m_{h_{125}} = 125.09 \; \mbox{GeV}
\eeq
by $h_{125}$. In the following, we exclude all parameter configurations where
this Higgs signal is built up by multiple resonances by demanding
the mass window $m_{h_{125}} \pm 5$~GeV to be free of any Higgs bosons except for
$h_{125}$.  Furthermore, we do not include electroweak
corrections in the parameter scans nor in the analysis, as they are
not (entirely) available for all observables and cannot be taken over from
the SM. \s

We check for the theoretical constraints on the N2HDM at tree level as
described in section \ref{sec:constraints}. Tree-level
perturbative unitarity is verified by using
Eqs.~(\ref{eq:ev1})-(\ref{eq:ev5}). This method yields a shorter
run-time than the model-independent numeric check implemented in {\tt
  ScannerS}. We checked that both methods lead to the same
results. Equations (\ref{eq:omega1}) and (\ref{eq:omega2})
are used to guarantee that the potential is bounded from below. The
vacuum state found by {\tt ScannerS} is required to be the global
minimum, otherwise it is rejected. As described in \ref{sec:global},
the check is performed by comparing
the value of the scalar potential at the {\tt ScannerS} vacuum with
the values at all of the stationary points. \s

Many of the experimental constraints applied on the 2HDM
also hold for the N2HDM. The constraints on $R_b$
\cite{Haber:1999zh,Deschamps:2009rh}
and $B \to X_s \gamma$
\cite{Deschamps:2009rh,Mahmoudi:2009zx,Hermann:2012fc,Misiak:2015xwa}
are only sensitive to the charged Higgs boson
so that the 2HDM calculation and the resulting $2\sigma$ exclusion bounds in the
$m_{H^\pm} - t_\beta$ plane can be taken over to the N2HDM. Note that
the latest calculation \cite{Misiak:2015xwa} enforces
\beq
m_{H^\pm} > 480 \mbox{ GeV}
\eeq
in the type II and lepton specific 2HDM. In the type I model on the
other hand the bound is much weaker and more strongly
  dependent on $\tan\beta$.
The oblique parameters $S$, $T$ and $U$ are calculated with the general
formulae in \cite{Grimus:2007if,Grimus:2008nb}, and $2\sigma$ compatibility
with the SM fit~\cite{Baak:2014ora} including the full correlations is demanded. \s

The N2HDM must comply with the LHC Higgs data. This requires one
scalar state to match the observed signal rates for a Higgs boson
of about 125 GeV. Furthermore, the remaining Higgs bosons must be
consistent with the exclusion bounds from the collider searches at
Tevatron, LEP and LHC, where the strongest constraints arise from the
LHC Run 1 data. {\tt ScannerS} provides an interface with {\tt
  HiggsBounds v4.3.1} \cite{Bechtle:2008jh,Bechtle:2011sb,Bechtle:2013wla}
which we use to check for agreement with
all $2\sigma$ exclusion limits from LEP, Tevatron and LHC Higgs searches.
The required input for {\tt HiggsBounds} are the cross
section ratios (relative to a SM Higgs boson of the same mass) of the
different production modes, the branching ratios and
the total widths for all scalars. We compute the latter two with the
program {\tt N2HDECAY}. The production cross sections through gluon
fusion (ggF) and $b$-quark fusion (bbF) are
obtained at next-to-next-to-leading order (NNLO) QCD from {\tt SusHi
  v1.6.0} \cite{Harlander:2012pb,Harlander:2016hcx} which is interfaced with {\tt
  ScannerS}.\footnote{{\tt SusHi} computes the cross sections for the
  2HDM. As the only change of the N2HDM with respect to the 2HDM
  besides an additional CP-even Higgs boson are the values of the
  effective couplings to the fermions {\tt SusHi} can also be used to compute
  the cross sections for the N2HDM
  Higgs bosons.}
The remaining cross section ratios at leading order (and
also at higher order in QCD) are given by the effective couplings
squared. For example for the production in association with a fermion
pair it is $c(H_i f\bar{f})^2$ (Table~\ref{tab:yukcoup}),
and for the gauge boson mediated cross sections (vector boson
fusion and associated production with a vector boson) it is
$c(H_i VV)^2$ (\eqref{eq:n2hdmgaugecoup}). \s

Compatibility of the discovered Higgs signal with $h_{125}$ is checked by using
the individual signal strengths fit of
Ref.~\cite{Khachatryan:2016vau}. The needed
decay widths and branching ratios are taken from {\tt
  N2HDECAY}. The fermion initiated cross section normalized to the
SM,
\beq
\mu_F = \frac{\sigma_{\text{N2HDM}} (ggF)
  +\sigma_{\text{N2HDM}} (bbF) }{\sigma_{\text{SM}} (ggF)} \;,
\eeq
is obtained with the NNLO QCD cross sections taken from {\tt SusHi}.
In the normalization we neglect the bbF cross section, which in
the SM is very small compared to gluon fusion. The production through
vector boson fusion (VBF) or through associated production with a
vector boson (VH) normalized to the SM, $\mu_V$, is given by
\beq
\mu_V = \frac{\sigma_{\text{N2HDM}}
  (VBF)}{\sigma_{\text{SM}} (VBF)} =
\frac{\sigma_{\text{N2HDM}}
  (VH)}{\sigma_{\text{SM}} (VH)} = c^2 (H_i VV) \;,
\eeq
where $H_i$ is identified with $h_{125}$. The QCD
corrections to massive gauge boson-mediated production cross sections
cancel upon normalization to the SM.
The properties of the $h_{125}$ are checked against the six fit
values of
\beq
\frac{\mu_F}{\mu_V} \;, \quad \mu_F^{\gamma\gamma} \;, \quad
\mu_F^{ZZ} \;, \quad \mu_F^{WW} \;, \quad  \mu_F^{\tau\tau} \;, \quad
\mu_F^{bb} \;,
\eeq
given in \cite{Khachatryan:2016vau}, with $\mu_F^{xx}$ defined as
\beq
\mu_F^{xx} = \mu_F \, \frac{\mbox{BR}_{\text{N2HDM}} (H_i \to
  xx)}{\mbox{BR}_{\text{SM}} (H_i \to xx)} \;.
\eeq
For $H_i \equiv h_{125}$ we require agreement with the fit results of
\cite{Khachatryan:2016vau} at the $2\times 1\sigma$ level. \s

In the numerical analysis we will show results for type I and type II
N2HDM models. For the scan with the input parameters from
Eq.~(\ref{eq:n2hdminputpars}) we fix $v$ to the SM value and choose
$t_\beta$ in the range
\beq
0.25 \le t_\beta \le 35 \;. \label{eq:tbscann2hdm}
\eeq
As the lower bound on $t_\beta$ from the $R_b$ measurement is stronger
than the lower bound in Eq.~(\ref{eq:tbscann2hdm}), the latter has no
influence on the physical parameter points. We transform the
mixing matrix generated by {\tt ScannerS} to the parametrisation of
Eq.~(\ref{eq:mixingmatrix}) such that the mixing angles are allowed to vary in the ranges
\beq
- \frac{\pi}{2} \le \alpha_{1,2,3} < \frac{\pi}{2} \;.
\eeq
We identify one of the neutral Higgs bosons $H_i$ with $h_{125}$ and
allow the remaining neutral Higgs bosons to have masses within
\beq
30 \mbox{ GeV } \le m_{H_i \ne h_{125}} , \; m_A \le 1 \mbox{ TeV} \;.
\eeq
In the type II model, the charged Higgs mass is chosen in the range
\beq
480 \mbox{ GeV } \le m_{H^\pm} < 1 \mbox{ TeV } \;,
\eeq
while we choose
\beq
80 \mbox{ GeV } \le m_{H^\pm} < 1 \mbox{ TeV }
\eeq
in type I. The singlet VEV $v_S$ is generated in the interval
\beq
1 \mbox{ GeV } \le v_S \le 1.5 \mbox{ TeV} \;.
\eeq
The value of $m_{12}^2$ is chosen as
\beq
0 \mbox{ GeV}^2 \le m_{12}^2  < 500 000 \mbox{ GeV}^2 \;.
\eeq
The condition $m_{12}^2> 0$ is found to be necessary for
the minimum to be the global minimum of the scalar potential.

\section{Phenomenological Analysis \label{sec:pheno}}
We start with the investigation of the parameter distributions and from
now on denote the lighter of the two non-$h_{125}$ CP-even Higgs bosons
by $H_\downarrow$ and the heavier one by $H_\uparrow$. \s

The inspection of the mass distributions resulting from our scan for
type II shows that the masses can  take all values between 30 GeV and
1 TeV. Furthermore, we note that it is possible that both $H_\downarrow$ and
$H_\uparrow$ are lighter than $h_{125}$ and also that $H_\downarrow$
and $A$ have masses below 125
GeV. Due to the lower bound $m_{H^\pm} \ge 480$~GeV and the
constraints from the EWPT, which force at least one of the non-SM-like
neutral Higgs bosons to have mass close to the charged Higgs mass,
there is no scenario with all non-SM-like neutral Higgs masses
below 125 GeV. \s

In type I, overall we have lighter Higgs spectra because of the much
weaker lower bound on the charged Higgs mass of $m_{H^\pm} \gsim
80$~GeV. Consequently, here we can also have situations where
$h_{125}$ is the heaviest Higgs boson in the spectrum.

\subsection{The wrong-sign Yukawa Coupling
  Regime \label{sec:wrongsign}}
\begin{figure}[t!]
\begin{center}
\includegraphics[width=8.5cm]{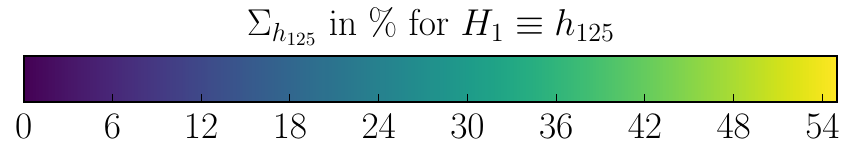}\\
\includegraphics[width=7.8cm,trim = 0mm 14mm 1mm
2mm,]{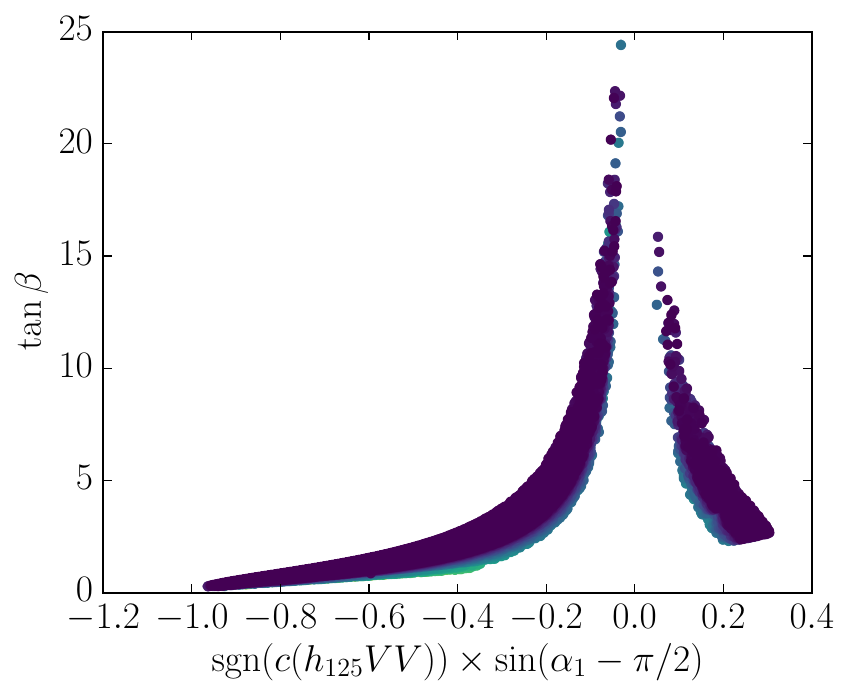}
\includegraphics[width=7.8cm,trim = 0mm 14mm 1mm
2mm,]{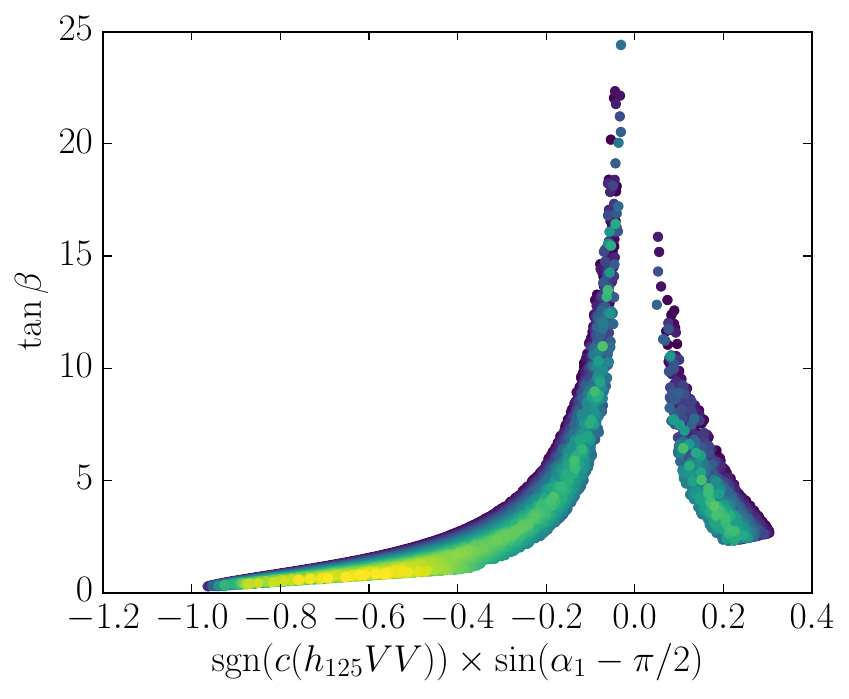}
\end{center}
\vspace*{0.2cm}
\caption{The wrong sign limit in the N2HDM as a function of the
  singlet admixture. The SM-like Higgs boson
  is given by $h_{125} \equiv H_1$. \label{fig:ws}}
\end{figure}

\begin{figure}[t!]
\begin{center}
\includegraphics[width=9cm,trim = 0mm 15mm 1mm
3mm,]{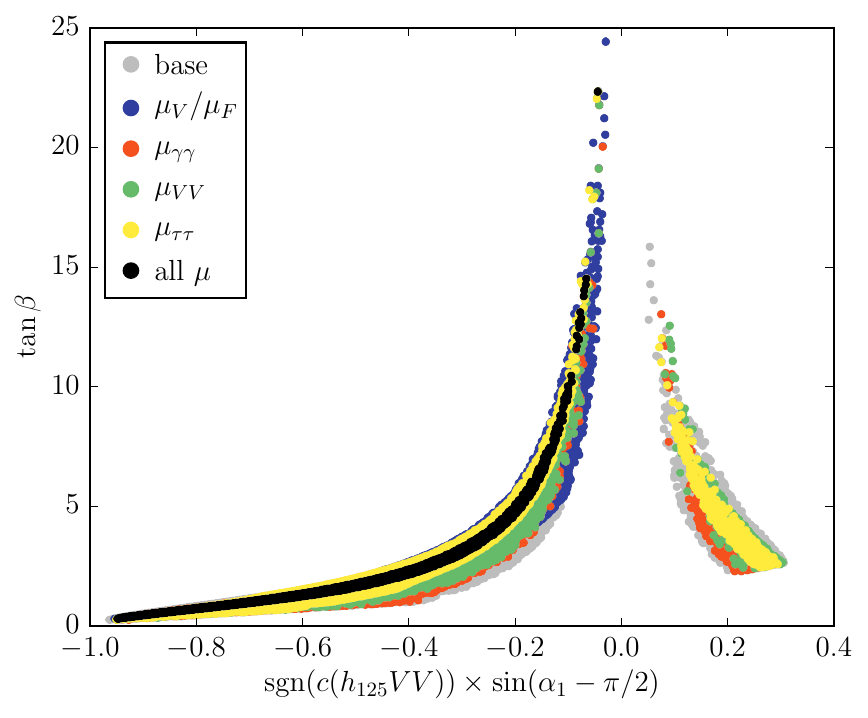}
\end{center}
\vspace*{0.3cm}
\caption{Constraining the N2HDM parameter space with $h_{125}=H_1$;
  grey: all scan
  points respecting the experimental and theoretical constraints;
  remaining colours: additional constraints of $0.95 \le \mu_V/\mu_F \le
  1.05$ (blue) and $0.95 \le \mu_{XX} \le
  1.05$ with $X=\gamma$ (orange),
  $X=V=Z$ (green), $X=\tau$ (yellow) and for all
  $\mu_{XX}$ and $\mu_V/\mu_F$ (black).
\label{fig:wsconstraint}}
\end{figure}

The wrong-sign Yukawa couplings regime, which was discussed in
\cite{Ferreira:2014naa,Ferreira:2014dya,Fontes:2014tga} for
the CP-conserving 2HDM, is the parameter
region where the coupling of the $h_{125}$ to the massive gauge bosons
is of opposite sign with respect to the coupling to fermions, $c(h_{125}
f \bar{f})$. This region is in contrast to the SM case
where both couplings have the same sign, which can have interesting
phenomenological consequences such as the non-decoupling of heavy particles
\cite{Ferreira:2014naa,Krause:2016xku}. This region is not excluded by
the experimental data for a
type II model with an opposite sign Yukawa coupling to down-type fermions. For the coupling to up-type fermions this wrong-sign
scenario is only realized for $\tan\beta <1$ which is excluded. In the
type I N2HDM the Yukawa couplings to up- and down-type fermions are the
same so that the wrong-sign coupling regime cannot be realized for any
of the quark types, as it requires $\tan\beta <1$. We investigate the
extent to which this scenario can be realized in the N2HDM type II. In
the 2HDM the wrong-sign regime is obtained for parameter values where
$\sin \alpha > 0$, while the
correct-sign regime is realized in the opposite case. In order to
match the 2HDM description, {\it cf.}~Eq.~(\ref{eq:2hdmlimit}), we
use the condition
\beq
\mbox{sgn} (c(h_{125} VV)) \cdot \sin (\alpha_1 - \tfrac{\pi}{2}) > 0
\eeq
for the wrong-sign limit. Figure~\ref{fig:ws} displays $\tan\beta$
versus $\mbox{sgn} (c(h_{125} VV)) \cdot \sin (\alpha_1 -
\frac{\pi}{2})$ for all parameter points from our N2HDM type II scan
that survive the imposed constraints and where the SM-like Higgs is
given by $H_1 = h_{125}$.\footnote{Quantities which explicitly involve
  the mixing angles $\alpha_i$ are inherently dependent on the mass
  ordering. For this reason we only consider the most frequent case
  $h_{125}=H_1$ in Figs. \ref{fig:ws} and \ref{fig:wsconstraint}.}
The colour code quantifies the
singlet admixture $\Sigma_{h_{125}}$ of the SM-like Higgs boson
$h_{125}$. We define the singlet admixture $\Sigma_{h_{125}}$ through
\beq
\Sigma_{h_{125}} \equiv |R_{h_{\text{SM}},3}|^2 \;,
\eeq
{\it i.e.}~the absolute value squared of the mixing matrix element
describing the mixing of the singlet field with the SM-like Higgs state.
In the right plot we have inverted the colour ordering.
In the left panel of Fig.~\ref{fig:ws} we observe a
large number of points in the 2HDM limit, {\it i.e.}~with small
singlet admixture. Such points are distributed in two branches with a
shape that agrees with the 2HDM. This allows to verify the identification of the left
branch with the correct-sign regime and of the right branch with the
wrong-sign regime. The inverted colour ordering in Fig.~\ref{fig:ws}
(right) allows us to investigate the repartition of the singlet
admixture over the two limiting cases. Overall we see that the singlet
admixture can be considerable. In the wrong-sign regime
it reaches up to about 30\% while in the correct-sign
regime it can even be as large as 55\%. The points with the largest singlet
admixture can be found for small values of $\tan\beta$.
In the following analysis we will comment further on the interesting
phenomenology of the wrong-sign regime. \s

\begin{figure}[t!]
\vspace*{0.4cm}
\begin{center}
\includegraphics[width=7.8cm,trim = 0mm 15mm 1mm
3mm,]{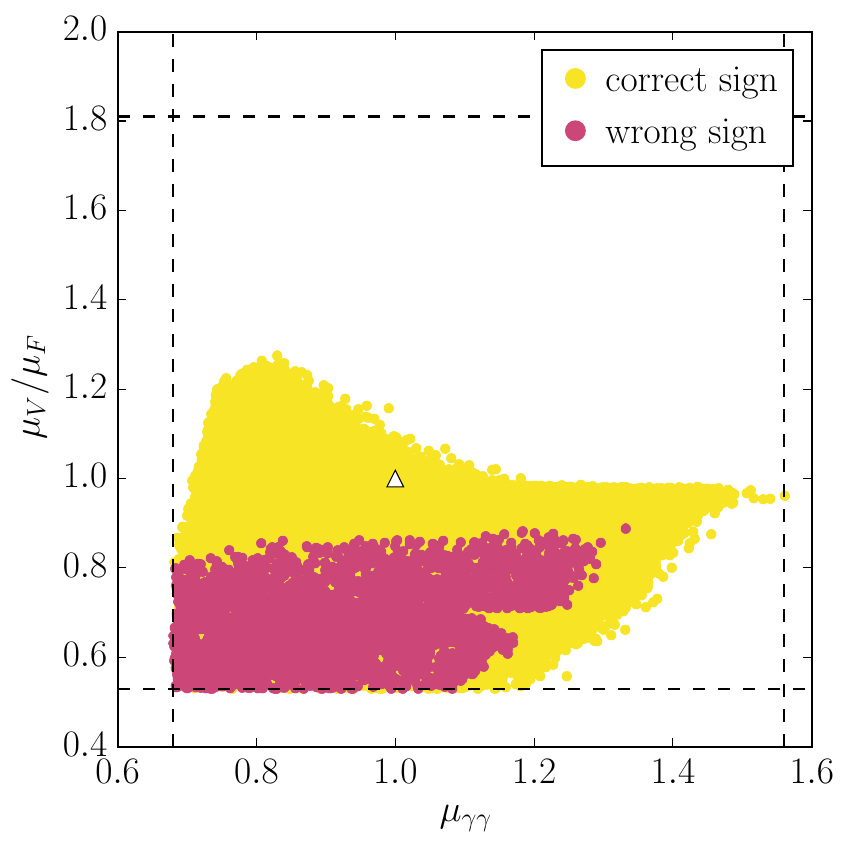}
\includegraphics[width=7.8cm,trim = 0mm 15mm 1mm
3mm,]{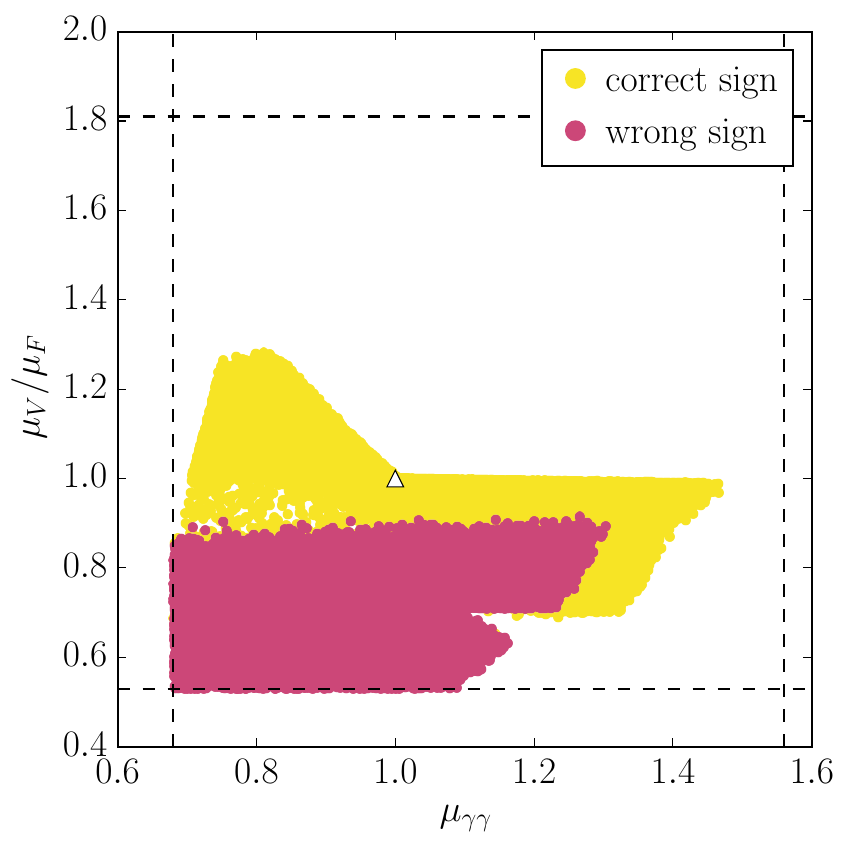}
\end{center}
\caption{The correct-sign (yellow) and the wrong-sign regime (pink)
  in the $\mu_V/\mu_F$-$\mu_{\gamma\gamma}$ plane for the N2HDM (left)
  and the 2HDM (right). The $h_{125}$ can be any of the $H_i$ (N2HDM) and
  $h_{125} \equiv h$ or $H$ (2HDM). The white triangle denotes the SM
  point. The dashed lines are the current limits on the $\mu$
  values \label{fig:wsmugmuvof}}
\end{figure}

\noindent
{\bf Constraining the wrong-sign regime}
An important question to ask is to which extent will the collection of
more precise data, obtained at the LHC Run II and in the
high-luminosity phase, be able to
constrain the N2HDM parameter space and in particular the wrong-sign
regime. In Fig.~\ref{fig:wsconstraint} we show again the allowed region in
the $\tan\beta$ versus
$\mbox{sgn}(c(h_{125}VV))\cdot\sin(\alpha_1-\pi/2)$ projection of the
parameter space, where the grey points respect all theoretical and
experimental constraints, in
particular the reduced signal strengths from the six-parameter fit of
Ref.~\cite{Khachatryan:2016vau} for the SM-like Higgs (here $H_1\equiv
h_{125}$). We then
successively constrain the $\mu$-values further, by assuming that
future more precise measurements are able to achieve a precision of
5\%, with a central value of~1. Hence
\beq
\begin{array}{lll}
0.95 \le \mu_{XX} \le 1.05 & \mbox{for} & X=\gamma \mbox{ (orange)}, V=Z
\mbox{ (green)}, \tau \mbox{ (yellow)} \\
0.95 \le \mu_V/\mu_F \le 1.05 & & \mbox{(blue)} \\
0.95 \le \mbox{ all } \le 1.05 & & \mbox{(black)}
\end{array}
\label{eq:colorcode}
\eeq
The plot shows that the correct-sign regime given by the left branch
is most strongly constrained by the $\tau\tau$ final state. While the
wrong-sign limit is also very sensitive to this observable its
compatibility with the data is fundamentally different as there are no
black nor blue points in the right branch. This behaviour is
represented in different form in Fig.~\ref{fig:wsmugmuvof}, in which
we have plotted $\mu_V/\mu_F$ versus $\mu_{\gamma\gamma}$.
The depicted yellow areas show the points from the parameter scan
compatible with all constraints in the correct-sign regime, for the
N2HDM (left plot) and, for comparison, also for the 2HDM
(right).\footnote{The constraints
  applied in the 2HDM are the same as for the N2HDM, as well as the scan
ranges (which are identical up to the additional mixing angles and masses
in the N2HDM).} The pink points represent
the wrong-sign regime. Here and from now on
  $h_{125}$ can be any of the CP-even, neutral Higgs bosons ({\it
    i.e.} any $H_i$ in the N2HDM and $h$ or $H$ in the 2HDM).
The white triangle denotes the SM result.
The yellow and pink regions completely overlap in
the N2HDM in contrast to the 2HDM. Here we have less parameters and
hence less freedom to reach compatibility with the constraints than in
the N2HDM so that the yellow area is more restricted.\footnote{The
  upper two limiting lines of the correct-sign regime
in the 2HDM are due to the unitarity of the mixing matrix, whose
elements enter the Higgs couplings.}
The allowed area of the wrong-sign regime, on the other hand, is the
same in both models. The reason is that in the N2HDM the singlet
admixture can at most reduce the Higgs couplings to SM particles and
hence the $\mu$-values. As the wrong-sign regime in the 2HDM is
already touching the lower bounds in the presented $\mu$-values the
N2HDM cannot add anything new to this region.
From these figures we immediately infer that in the wrong-sign regime the ratio
$\mu_V/\mu_F$ cannot reach~1, which explains the missing blue and
black points in Fig.~\ref{fig:wsconstraint}. The measurement of
$\mu_V/\mu_F$ is hence a powerful observable to constrain the
wrong-sign regime with values of $\mu_V/\mu_F \gsim 0.9$, excluding
this scenario. We note that the pink region in
Fig.~\ref{fig:wsmugmuvof} is rather insensitive to an
increase in the precision of $\mu_{VV}$ to 5\% around 1. \s
\begin{figure}[t!]
\begin{center}
\includegraphics[width=7.8cm,trim = 0mm 15mm 1mm
3mm,]{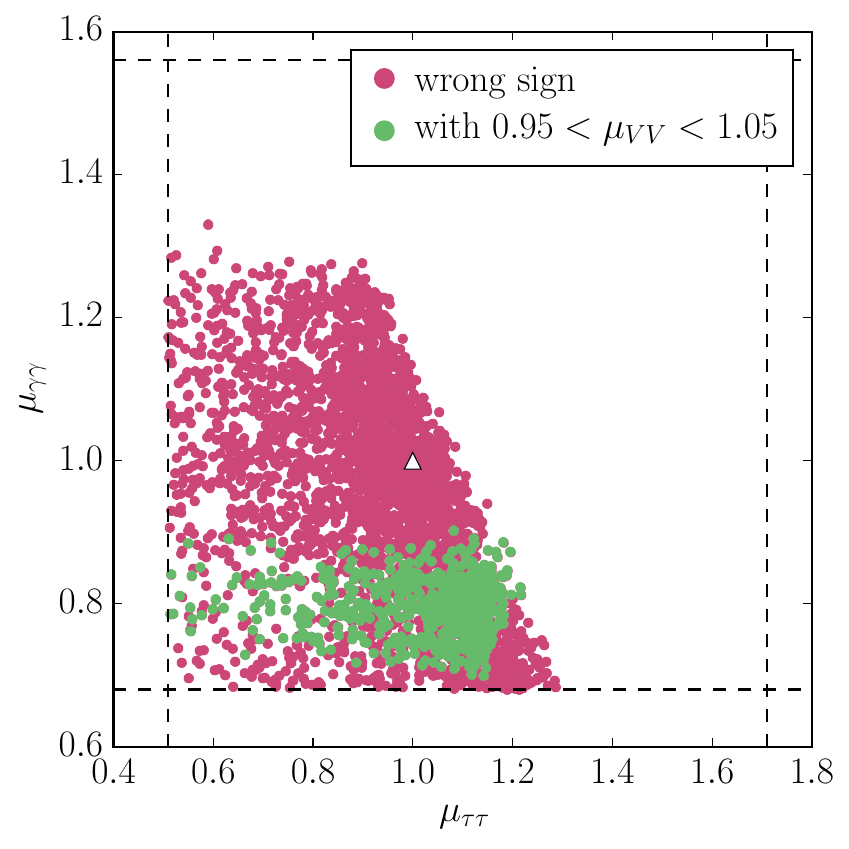}
\includegraphics[width=7.8cm,trim = 0mm 15mm 1mm
3mm,]{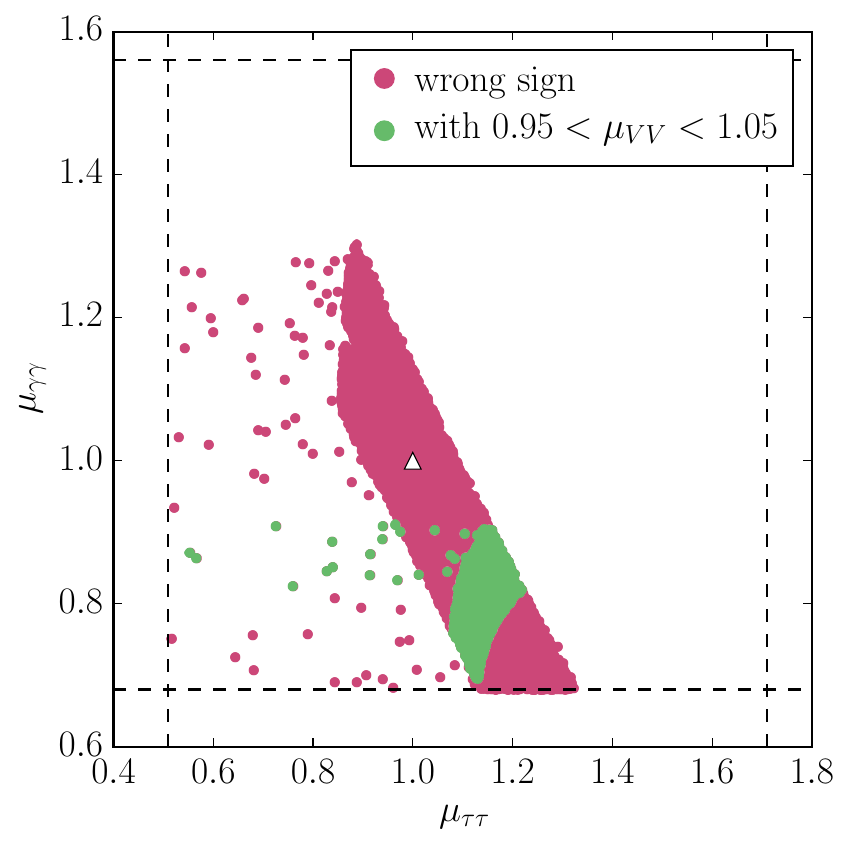}
\end{center}
\vspace*{0.2cm}
\caption{The wrong-sign regime
  in the $\mu_{\gamma\gamma}$-$\mu_{\tau\tau}$ plane for the N2HDM (left)
  and the 2HDM (right), where $h_{125}$ can be any of the $H_i$ (N2HDM) and
  $h_{125} \equiv h$ (2HDM). The green points restrict the allowed
  area (pink) further by imposing a higher precision in $\mu_{VV}$,
  $0.95 \le \mu_{VV} \le 1.05$. The dashed lines are the current limits on the
  $\mu$-values. \label{fig:wsmugmutautau}}
\end{figure}

In Fig.~\ref{fig:wsmugmutautau} the pink points show the wrong-sign
regime in the $\mu_{\gamma\gamma}$ versus $\mu_{\tau\tau}$ plane. The left panel is for the N2HDM, with $h_{125}$ being any of the $H_i$, whereas the right panel is for the 2HDM, with\footnote{The plot for
  $h_{125} \equiv H$ looks similar.} $h_{125} \equiv h$. We show on
top, in green, a further restriction of the sample assuming that
future measurements can constrain the
$\mu_{VV}$, $V=Z$, value to 5\% around the SM value. Again we observe that the
2HDM area, with a smaller number of parameters, is more constrained than the N2HDM, with the upper bounds in both models
being about the same. More importantly, we note that the increase in
the precision of
$\mu_{VV}$ forces the reduced photonic rate to be below about
0.9.\footnote{This was already observed for the 2HDM in
  \cite{Ferreira:2014naa,Ferreira:2014dya}.}
Thus, the the wrong-sign regime can be excluded by increasing the
precision in the $\mu_{VV}$ measurement and observing
$\mu_{\gamma\gamma} \gsim 0.9$.
The outliers in Fig.~\ref{fig:wsmugmutautau}
are points where $h_{125}$ has a substantial decay width into an (off-shell) pair of light Higgs bosons.
The resulting increase of the total width reduces the branching ratio into $\tau\tau$ and thus $\mu_{\tau\tau}$.
In the majority of the scenarios the light Higgs boson is the pseudoscalar $A$. If, however, $h_{125}$ is not the lightest CP-even neutral Higgs boson, decay widths of $h_{125}\rightarrow H_1 H_1$ ($h_{125}\rightarrow h h$ in the 2HDM) can also be substantial.

\subsection{Phenomenology of the $h_{125}$ Singlet
  Admixture \label{sec:singletadm}}

\begin{figure}[t!]
\begin{center}
\includegraphics[width=8.5cm]{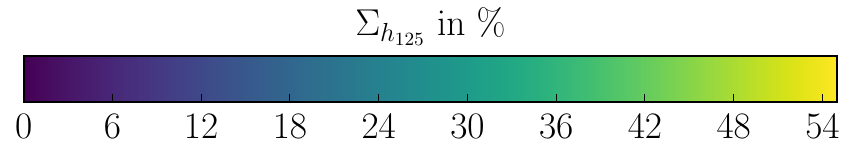}\\
\includegraphics[width=7.8cm,trim = 0mm 15mm 1mm
3mm,]{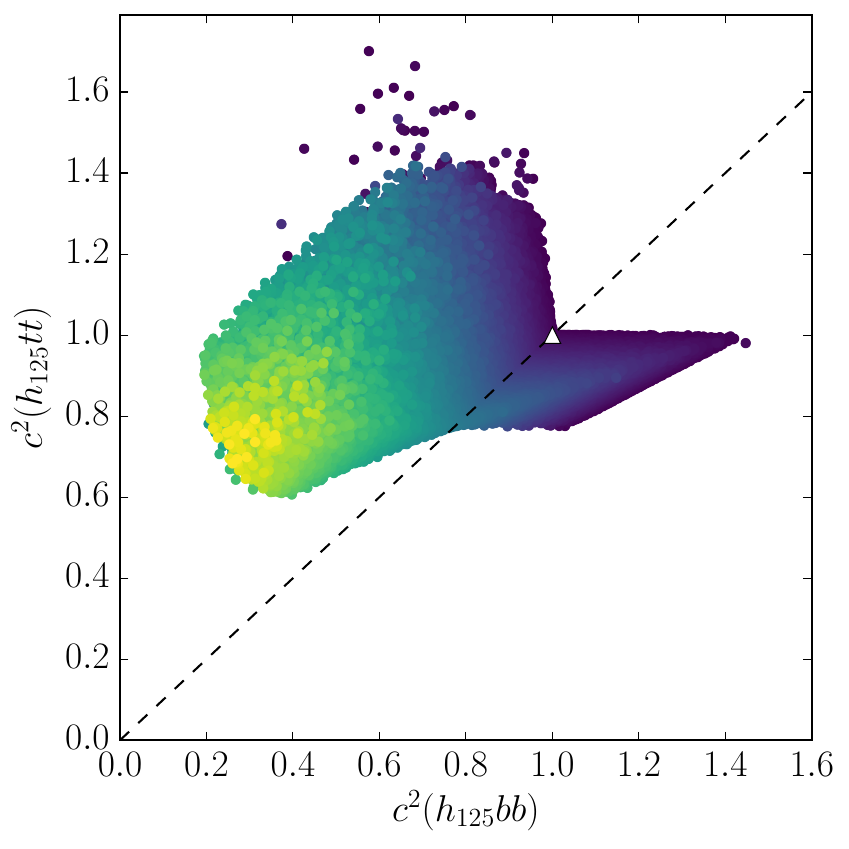}
\includegraphics[width=7.8cm,trim = 0mm 15mm 1mm 3mm,]{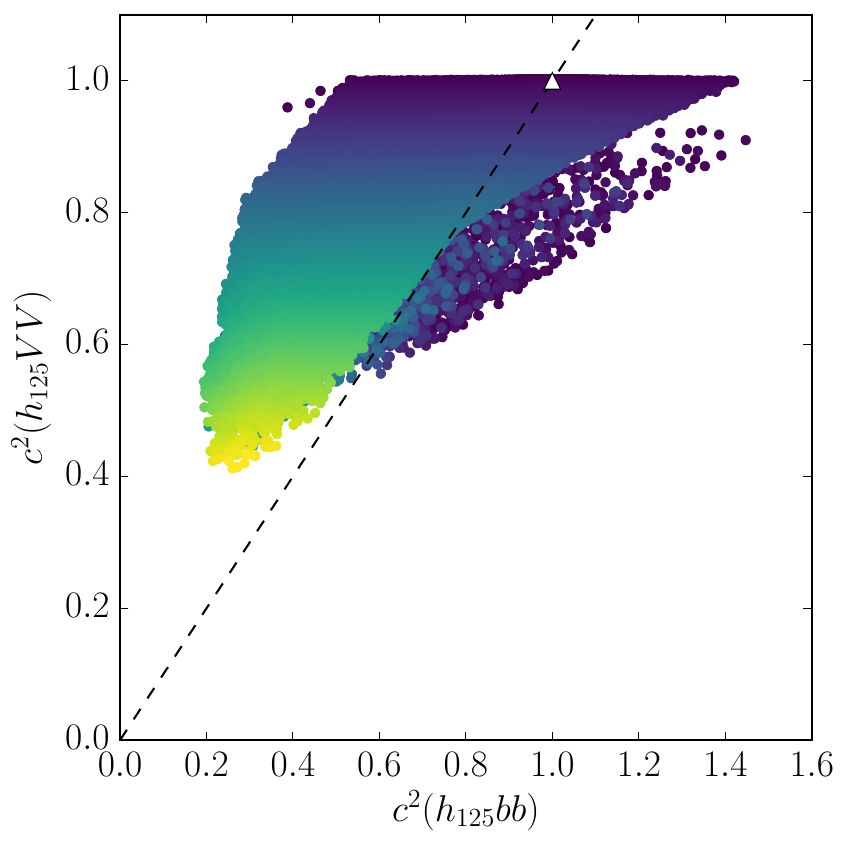}
\end{center}
\vspace*{0.2cm}
\caption{Type II: Singlet admixture $\Sigma_{h_{125}}$ as a function of the
  effective couplings squared. The white triangle denotes the SM
  value. The dashed line respresents equal scaling of the
  couplings. \label{fig:coupvalues}}
\end{figure}

\begin{figure}[b!]
\begin{center}
\includegraphics[width=8.5cm]{colorbar_T2}\\
\includegraphics[width=7.8cm,trim = 0mm 15mm 1mm
3mm,]{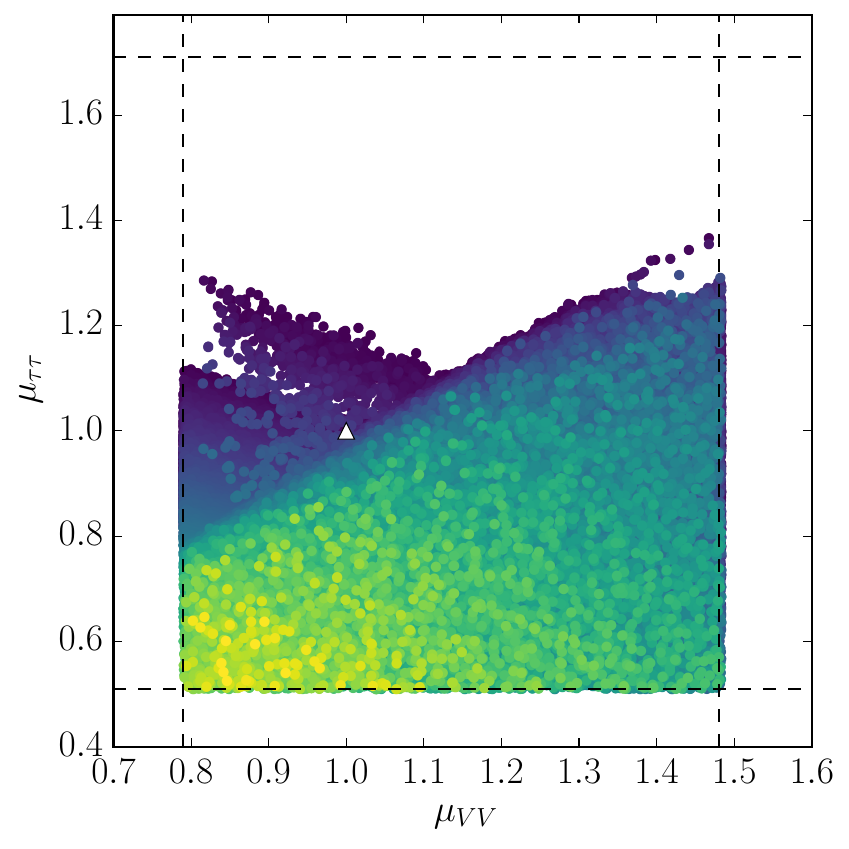}
\includegraphics[width=7.8cm,trim = 0mm 15mm 1mm 3mm,]{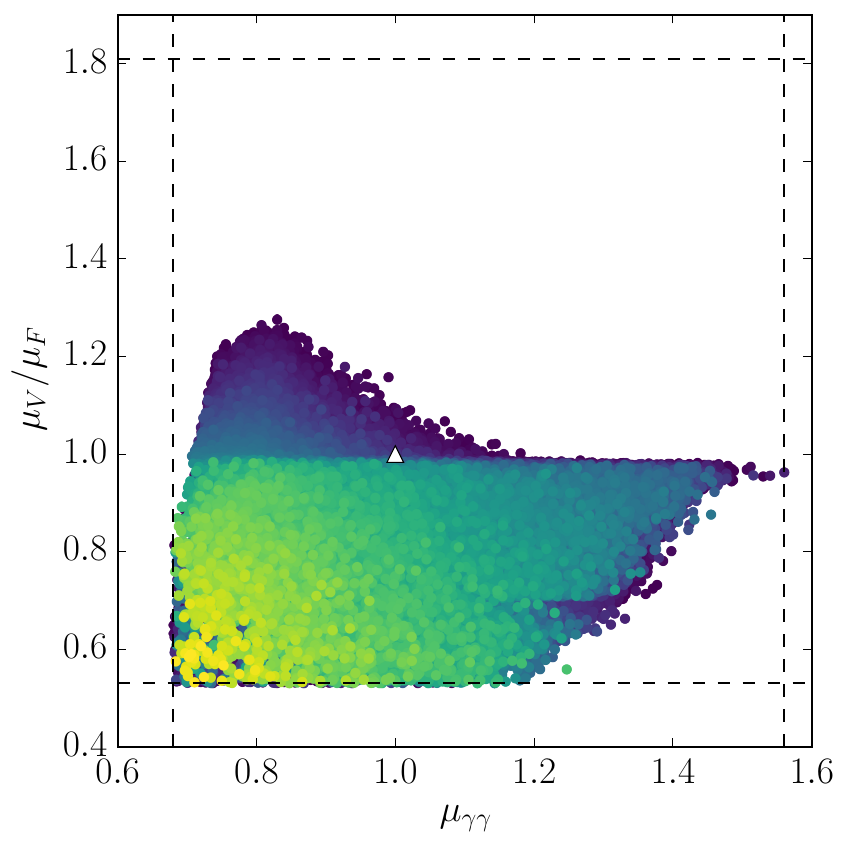}
\end{center}
\vspace*{0.2cm}
\caption{Type II: Singlet admixture $\Sigma_{h_{125}}$ as a function of the
  most constraining signal strengths. The white triangle denotes the
  SM value.\label{fig:muvalues}}
\end{figure}

\begin{figure}[t!]
\begin{center}
\includegraphics[width=8.5cm]{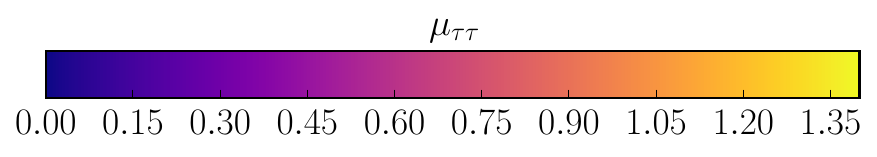}\\
\includegraphics[width=7.8cm,trim = 0mm 15mm 1mm
3mm,]{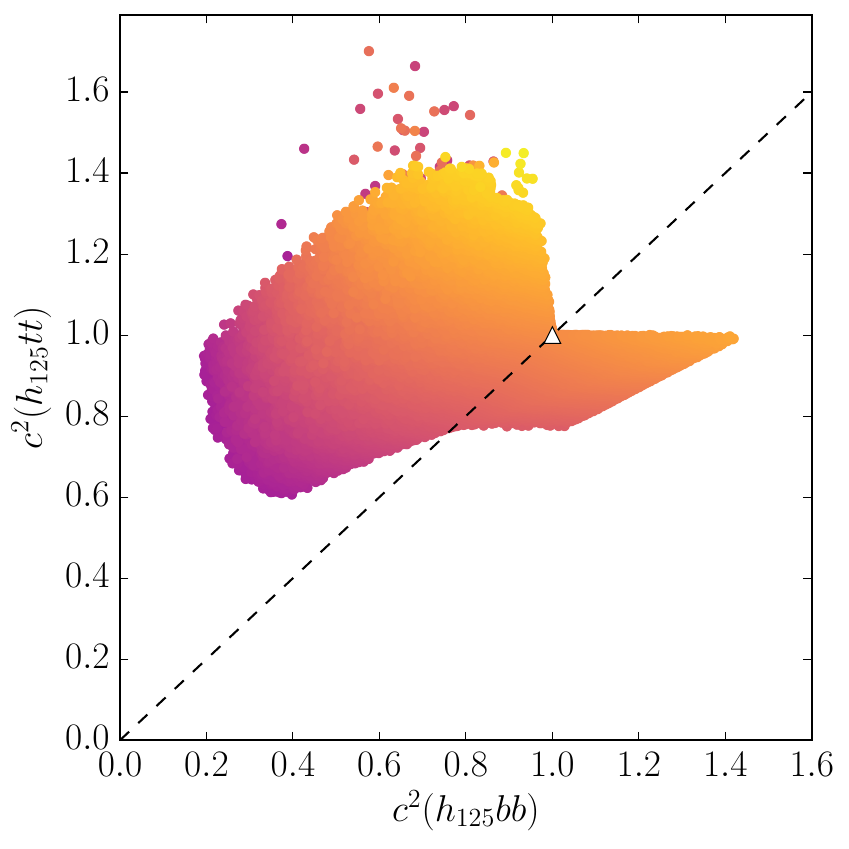}
\includegraphics[width=7.8cm,trim = 0mm 15mm 1mm 3mm,]{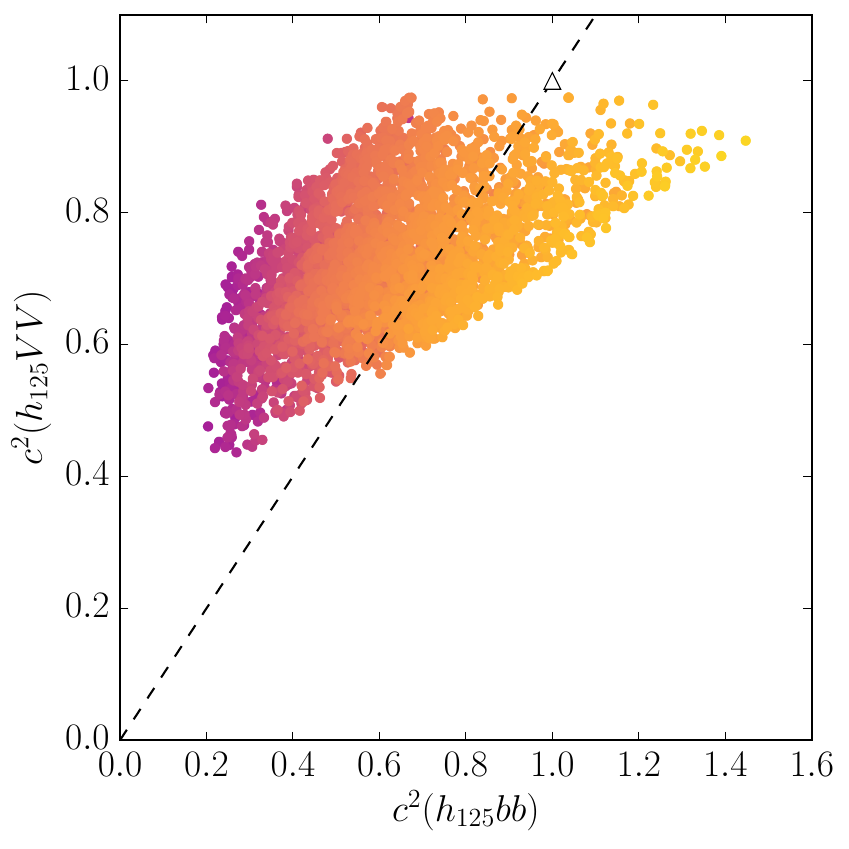}
\end{center}
\vspace*{0.2cm}
\caption{Type II: The $\tau\tau$ signal strength $\mu_{\tau\tau}$ as
  a function of the effective couplings squared. The white triangle denotes the SM
  value. The left plot shows only points in the correct-sign regime, the
  right plot only points in the wrong-sign regime.
The dashed line respresents equal scaling of the
  couplings. \label{fig:mutautaucoupvalues}}
\end{figure}

\noindent
{\bf Type II N2HDM:}
The large number of parameters in the N2HDM allows for considerably
non-standard properties in the phenomenology of the SM-like Higgs boson. In
particular, in type II, which we discuss first, significant singlet
admixtures of up to 55\% are still compatible with
the LHC Higgs data. This can be inferred from
Figs.~\ref{fig:coupvalues} and~\ref{fig:muvalues}.
Figure~\ref{fig:coupvalues} displays the correlation between pairs of
the effective couplings squared of $h_{125}$ to the SM particles. The
left plot shows the coupling to top quarks versus the coupling to
bottom quarks, and the right plot shows the
coupling to the massive gauge bosons versus the coupling to bottom quarks.
We remind that $h_{125}$ can be any of the
neutral CP-even Higgs bosons. The
influence of the singlet admixture is quantified by the colour code.
In Fig.~\ref{fig:muvalues} (left) we show the
reduced signal rate in $\tau$ final states, $\mu_{\tau\tau}$, versus the
one into massive vector bosons, $\mu_{VV}$. In Fig.~\ref{fig:muvalues}
(right) we plot the ratio $\mu_V/\mu_F$ of the vector boson induced
production over the fermionic production, each normalized to the SM,
versus the photon final state signal strength $\mu_{\gamma\gamma}$.
The white triangle indicates the SM
values of the signal strengths. The dashed lines are the experimental
limits on the respective signal strengths.
They show that the N2HDM parameter space is constrained by the upper
and lower limits on $\mu_{VV}$ and $\mu_{\gamma\gamma}$ and the lower
limits on $\mu_{\tau\tau}$ and $\mu_V/\mu_F$, respectively. As can be
inferred from Fig.~\ref{fig:muvalues} (left) enhanced rates in the
$\tau$ final state are still allowed by the experimental data where
the largest enhancement is reached for small admixture, {\it
  i.e.}~in the 2HDM-like regions. The area of the enhanced
$\mu_{\tau\tau}$ can be divided in three regions that shall be
explained separately. The largest enhancement of up to 40\% is
obtained for simultaneously enhanced $\mu_{VV}$. The enhancement is
due to the production mechanism and corresponds to the enhanced
couplings to top quarks in Fig.~\ref{fig:coupvalues} (left) while the
involved decays remain SM-like.
This is confirmed by Fig.~\ref{fig:mutautaucoupvalues} (left), which shows
the value of $\mu_{\tau\tau}$ in the plane of the effective couplings
squared, $c^2 (h_{125} tt)$ and $c^2(h_{h125} bb)$, for parameter
points in the correct-sign regime. The
largest $\mu_{\tau\tau}$, given by the yellow points, are found for
large effective couplings to top quarks.
The 2HDM-like region in Fig.~\ref{fig:muvalues}, where we have enhanced
$\mu_{\tau\tau}$ values but reduced $\mu_{VV}$, is due to enhanced
effective couplings to $\tau$
leptons and $b$ quarks, {\it i.e.}~the spikes in
Fig.~\ref{fig:coupvalues} and Fig.~\ref{fig:mutautaucoupvalues} (left),
respectively, where $\mu_{\tau\tau}$ reaches values of up to about 1.4.
On the other hand, the effective coupling to gauge bosons
cannot exceed one, so that overall the branching ratio
into gauge boson is reduced in favor of $\mbox{BR}(h_{125} \to
\tau\tau)$.
Finally, the points with $\mu_{\tau\tau} >1$
and $\mu_{VV} \approx 1$ are located in the wrong-sign regime and have
reduced couplings to the gauge bosons. In Fig.~\ref{fig:coupvalues}
(right) these are the points below the dashed line and isolated from
the bulk of the points. Figure~\ref{fig:mutautaucoupvalues} (right) shows the
same coupling values squared, but now only for points in the
wrong-sign limit, where the area with enhanced $\mu_{\tau\tau}$ can
easily be identified. It is the resulting reduced decay width into
$VV$ that increases the branching ratio into $\tau\tau$ and thus
$\mu_{\tau\tau}$.
Note that in Fig.~\ref{fig:coupvalues} (left)
the sharp lines departing from the SM point to smaller and larger
values of $c^2(h_{125} bb)$ are again due to the unitarity of the
mixing matrix. \s

While the enhanced $\mu_{\tau\tau}$ is a feature of the 2HDM-like
regions, singlet admixtures as large as 55\% can be compatible
with the current data. Interestingly the best measured
quantities $\mu_{VV}$ and $\mu_{\gamma\gamma}$  are not the ones with the highest constraining power on the singlet admixtures. A value of $\mu_{VV} =1$ still
allows for $\Sigma_{h_{125}}$ of up to 50\%, and a value
of $\mu_{\gamma\gamma}=1$ can be compatible with $\Sigma_{h_{125}}$
values of up to about 40\%. A
measurement of $\mu_{\tau\tau} \approx 1$ on the other hand constrains
$\Sigma_{h_{125}}$ to be below about
25\%. And a measurement of $\mu_V / \mu_F \ge 1$ enforces
$\Sigma_{h_{125}} \lsim 20$\%. This behaviour can be understood by
inspecting the involved couplings individually. Overall, high singlet
admixtures induce reduced
couplings. However, Fig.~\ref{fig:coupvalues} shows that the effective
coupling to $b$-quarks is reduced more strongly than that to
top-quarks, where the latter dominates gluon fusion
production. Furthermore, among the physical points, a large singlet
admixture leads to the effective gauge coupling being larger than the
effective bottom coupling. This means
that the total width is reduced due to the strongly reduced
$\Gamma(h_{125} \to b\bar{b})$, which dominates the total width in the
SM case. In turn, this increases the $\mbox{BR} (h_{125} \to VV)$ and
$\mbox{BR} (h_{125} \to \gamma\gamma)$ strongly enough to balance the
reduced gluon fusion production cross section and the reduced partial
widths in these channels, so that overall SM-values of
  $\mu_{VV}$ and $\mu_{\gamma\gamma}$ can be compatible with large
  singlet admixtures. The decay width into $\tau$ leptons, in
contrast, gets rescaled by the same coupling as the $b\bar{b}$ channel
in the type II model, so that the $\mu_{\tau\tau}$ is reduced for
large singlet admixtures. Since with increasing singlet
admixture $c(h_{125} VV)$ is more strongly reduced than $c(h_{125}
tt)$ the value $\mu_V/\mu_F$ is reduced for non-zero singlet
admixtures. As the coupling to gauge bosons reaches at most~1, an
enhanced $\mu_V/\mu_F$ can only be due to a smaller coupling of
$h_{125}$ to $tt$. A value of $c(h_{125} VV)$ close to~1 with a simultaneously
reduced $c(h_{125}tt)$ is only possible for small singlet admixtures.
The corresponding bulge in Fig.~\ref{fig:muvalues}
(right) contains points for which
all other $\mu$-values are close to their lower experimental
boundaries. They are characterised by large couplings of the SM-like
Higgs to a charged Higgs pair, which enters the loop-induced Higgs decay into
photons and keeps $\mu_{\gamma\gamma}$ above its lower experimental boundary. \s

\begin{figure}[t!]
\begin{center}
\includegraphics[width=9cm,trim = 0mm 15mm 1mm 3mm,]{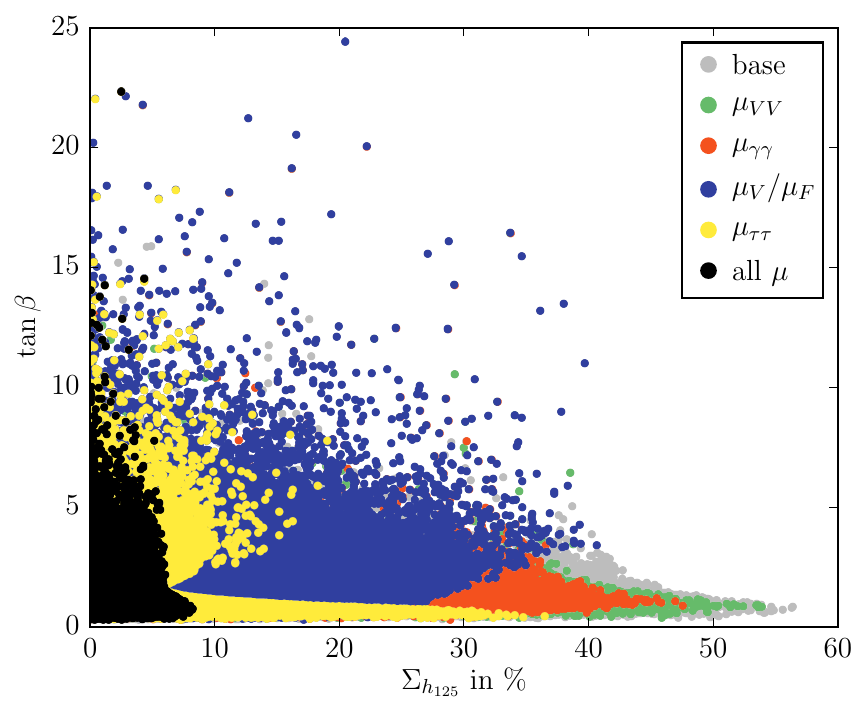}
\end{center}
\vspace*{0.2cm}
\caption{Type II: Singlet admixture $\Sigma_{h_{125}}$ as a function of
  $\tan\beta$ and the precision in the $\mu$-values.
  Grey points: all N2HDM scan parameter points; the remaining colors
  denote the $\mu$ values measured within 5\% of the SM
  reference. \label{fig:singadm}}
\end{figure}

We conclude our discussion of the type II N2HDM by displaying in
Fig.~\ref{fig:singadm} the
allowed N2HDM parameter
region in the $\tan\beta$ versus $\Sigma_{h_{125}}$ plane (grey points) and its
subsequent restriction by more precise measurements, with the same colour
code as defined in Eq.~(\ref{eq:colorcode}). It
reflects our findings that the singlet admixture is most powerfully
constrained by a precise measurement of $\mu_{\tau\tau}$ while being much less
sensitive to the remaining $\mu$-values. The restriction power depends
on the value of $\tan\beta$. For medium values of $\tan\beta$ a singlet
admixture of up to 20\% is still compatible with a 5\% measurement
of $\mu_{\tau\tau}$ whereas for small values of $\tan\beta$ even up to
37\% is allowed. Only the simultaneous measurement of all $\mu$-values
within 5\% around the SM value restricts $\Sigma$ to values below about 8\%. \s
\begin{figure}[t!]
\begin{center}
\includegraphics[width=8.5cm]{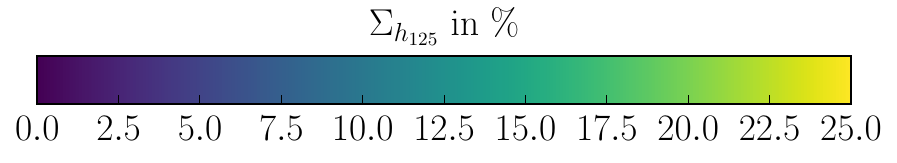}\\
\includegraphics[width=7.8cm,trim = 0mm 15mm 1mm
3mm,]{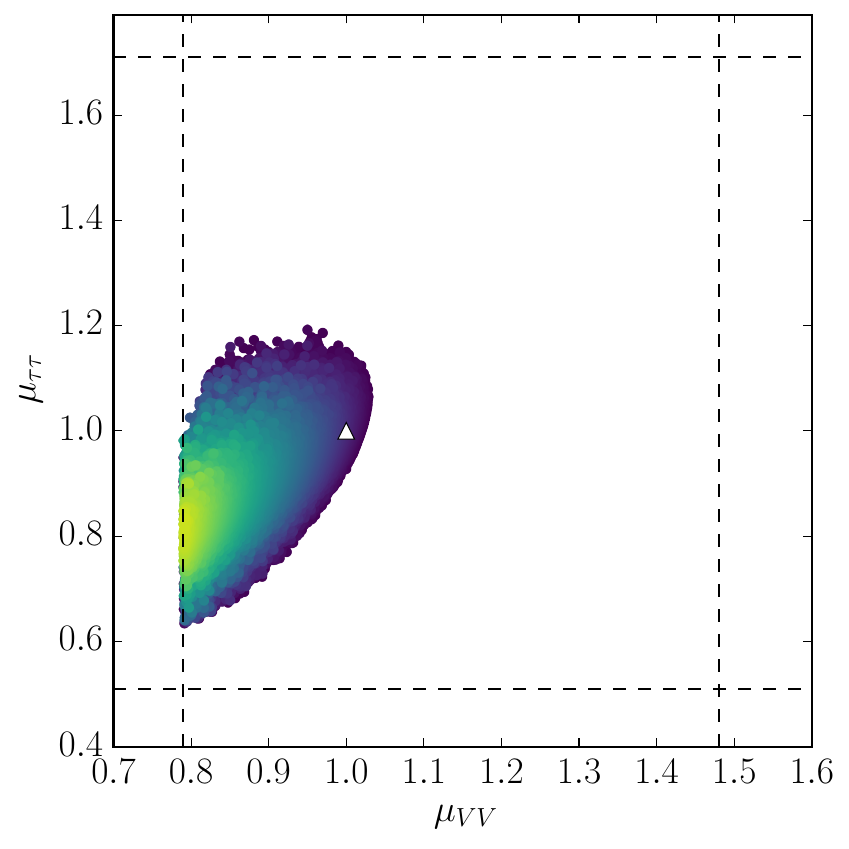}
\includegraphics[width=7.8cm,trim = 0mm 15mm 1mm
3mm,]{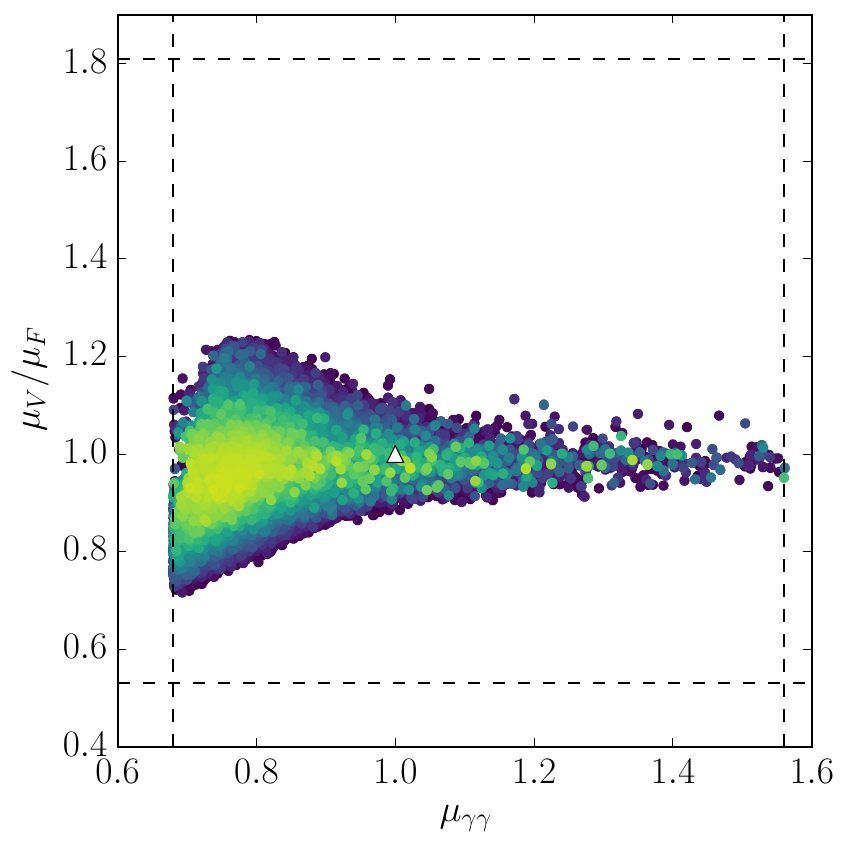}
\end{center}
\vspace*{0.2cm}
\caption{Type I: Singlet admixture $\Sigma_{h_{125}}$ as a function of the
  most constraining signal strengths. The white triangle denotes the
  SM value.\label{fig:t1muvalues}}
\end{figure}
\begin{figure}[b!]
\begin{center}
\includegraphics[width=9.5cm,trim = 0mm 15mm 1mm 3mm,]{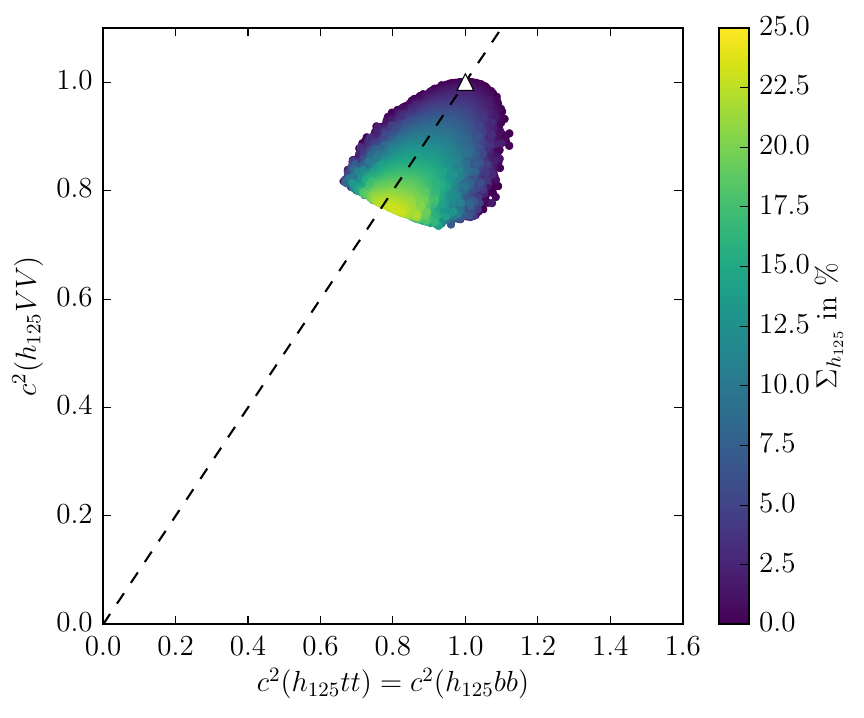}
\end{center}
\vspace*{0.2cm}
\caption{Type I: Singlet admixture $\Sigma_{h_{125}}$ as a function of the
  effective couplings squared. The white triangle denotes the SM
  value. The dashed line respresents equal scaling of the
  couplings. \label{fig:t1coupvalues}}
\end{figure}

\noindent
{\bf Type I N2HDM:}
We now turn to the discussion of the N2HDM type I. In this case the
doublet $\Phi_2$ couples to both up- and down-type
quarks. Consequently, any change in the coupling to the top quark is
induced also in the bottom quark coupling and vice versa. This
restricts the available freedom in the choice of the parameters of the
N2HDM so that, overall, less pronounced deviations from the SM or from the
pure 2HDM are expected. This can be inferred from
Figs.~\ref{fig:t1muvalues} and \ref{fig:t1coupvalues}, which show, respectively, the
singlet admixture of $h_{125}$ as a function of the reduced signal
strengths and the effective couplings. In constrast to
the type II N2HDM the singlet admixture can reach at most 25\%.
Figure~\ref{fig:t1muvalues} shows the most constraining
signal strenghts, {\it i.e.}~$\mu_{\tau\tau}$ versus $\mu_{VV}$ in the
left plot and $\mu_V/\mu_F$ versus $\mu_{\gamma\gamma}$ in the right
plot. From this figure it can be
inferred that the allowed areas in these two planes are more reduced
than in type II. While $\mu_{\gamma\gamma}$ is delimited by
the present LHC data, only the lower bound of $\mu_{VV}$ is
restricted due to the LHC fit value, while the upper $\mu_{VV}$ bound
and the boundaries of $\mu_{\tau\tau}$ and $\mu_V/\mu_F$ of the
allowed areas are already well below the restrictions set by the LHC
data. The highest singlet admixtures come with reduced
signal strengths while the ratio $\mu_V/\mu_F \approx 1$. This is in
accordance with Fig.~\ref{fig:t1coupvalues}, which shows the effective
couplings squared, $c(h_{125} VV)^2$ versus $c^2(h_{125} bb)$ ($=c^2 (h_{125} tt)$)
together with the singlet
admixture. Both effective couplings are reduced almost in parallel
with rising $\Sigma_{h_{125}}$ so that $\mu_V/\mu_F \sim c(h_{125}
VV)^2/c(h_{125} ff)^2$ is close to one for large singlet
admixtures. \s

\begin{figure}[t!]
\begin{center}
\includegraphics[width=9cm,trim = 0mm 15mm 1mm 3mm,]{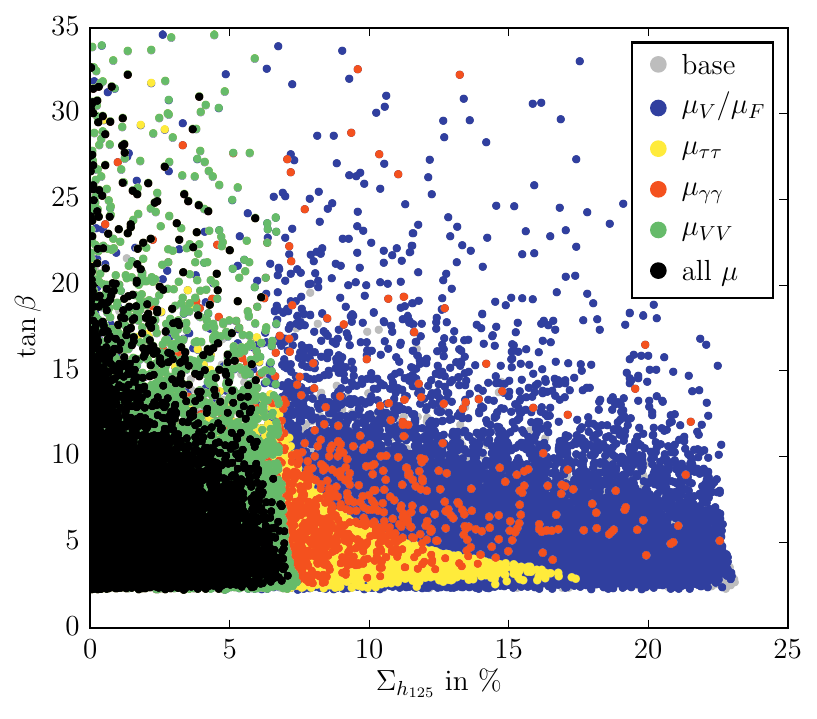}
\end{center}
\vspace*{0.2cm}
\caption{Type I: Singlet admixture $\Sigma_{h_{125}}$ as a function of
  $\tan\beta$ and the precision in the $\mu$-values.
  Grey points: all N2HDM scan parameter points; the remaining colors
  denote the $\mu$ values measured within 5\% of the SM
  reference. \label{fig:singadmt1}}
\end{figure}
The SM-like dark-blue boundary in Fig.~\ref{fig:t1muvalues} (left) with
enhanced $\mu_{\tau\tau}$ corresponds to the right boundary of the
area in Fig.~\ref{fig:t1coupvalues} where $c^2(h_{125} bb)$ is enhanced
and $c^2(h_{125} VV)$ is reduced. The simultaneous enhancement of the
Higgs coupling to the bottom quarks and $\tau$ leptons leaves the
branching ratio into $\tau$'s unchanged, but the enhanced Higgs
couplings to fermions enhance the main production mechanism so that
overall $\mu_{\tau\tau}$ is enhanced. Applying the same reasoning it
is clear that the dark-blue boundary with reduced $\mu_{\tau\tau}$
values corresponds to the left dark-blue boundary in
Fig.~\ref{fig:t1coupvalues}. This is also the area in
Fig.~\ref{fig:t1muvalues} (right) corresponding to
enhanced $\mu_V/\mu_F$ values for $\mu_{\gamma\gamma} \lsim 1.3$ in
the 2HDM-like region (blue area). The
reduced coupling to the bottom quarks reduces the dominating decay
into $b\bar{b}$. The stronger reduction of $c(h_{125} bb)$ as compared
to $c(h_{125} VV)$ overcompensates the reduced $\Gamma (h_{125} \to
VV)$ so that overall $\mbox{BR}(h_{125} \to VV)$ is enhanced and makes up for
the reduction of the production cross section due to the smaller
couplings to fermions. Simultaneously, this also ensures that
$\mu_V/\mu_F$ is enhanced. An analogous reasoning allows to identify
the dark blue lower $\mu_V/\mu_F$ values for $\mu_{\gamma\gamma}
\lsim 1.3$ with the right boundary in Fig.~\ref{fig:t1coupvalues}.  \s

Figure~\ref{fig:singadmt1} allows to analyse by which measurement the singlet
admixture can be most effectively constrained. The colour code has
been defined in Eq.~(\ref{eq:colorcode}). As can already be inferred
from Fig.~\ref{fig:t1muvalues} the singlet admixture is mostly
restricted by the precise measurement of $\mu_{ZZ}$,
down to about 7.5\% for a 5\% accuracy in $\mu_{ZZ}$. The simultaneous
measurement
of all $\mu$-values with a precision of 5\% hardly improves on this
constraint. When comparing with the type II case, we can conclude that
the structure of the Yukawa couplings decides which
final state in the Higgs data is the most effective one in
constraining the N2HDM, {\it i.e.}~its singlet admixture.

\section{Conclusions \label{sec:concl}}
In this paper we have investigated the N2HDM, which is based on
the CP-conserving 2HDM extended by a real scalar singlet field. It combines a
parameter space which is larger than the 2HDM with a greater freedom
in the choice of
the parameters (compared to singlet extended supersymmetric models as
{\it e.g.}~the next-to-minimal supersymmetric model). This allows for
an interesting phenomenology that is still compatible with the
experimental data. Thus the Higgs couplings can carry a substantial
singlet admixture. However, in order to be able to determine the
allowed parameter space and thereby perform meaningful phenomenological
analyses, the investigation of the constraints on the model had to be put on solid
ground. In this paper we have performed a thorough analysis of the
theoretical constraints on the N2HDM Higgs potential. First, we have collected from the literature the formulae for the N2HDM that test tree-level perturbative unitarity and stability of the vacuum. Moreover, for the first time, we
have presented a detailed analysis of all the stationary points of the
potential to obtain its global minimum. The model,
together with the theoretical constraints, has been implemented in {\tt
  ScannerS}. For the test of the experimental constraints the necessary
branching ratios were obtained with the new program {\tt N2HDECAY}. We
have written
this code based on {\tt HDECAY} to provide the N2HDM branching ratios
and total widths including the state-of-the-art higher order QCD
corrections and off-shell decays. With these preparatory works
completed we were then in the
position to subject the N2HDM to critical theoretical and experimental
scrutiny in the second part of the article. \s

Taking into account the theoretical and experimental constraints in the N2HDM, we
performed a parameter space scan and analysed the properties of the allowed regions. In the type II N2HDM
substantial singlet admixtures of up to 55\% were found to be compatible with the data. It turned out that the most precisely measured
quantities, $\mu_{VV}$ and $\mu_{\gamma\gamma}$, are not the most
effective ones in constraining the singlet admixture, but instead it is the
$\mu_{\tau\tau}$ observable. In the type I N2HDM, on the other hand, the parameter space is more constrained due to the universal fermion coupling to up- and
down-type quarks and hence the allowed singlet admixture remains below
about 25\%. In the future, in this case, the singlet admixture will be most strongly
constrained by the precise measurement of $\mu_{ZZ}$. \s

Like in the 2HDM we find a wrong-sign regime in the N2HDM. While
overall the allowed parameter space is larger in the N2HDM compared to
the 2HDM, the wrong-sign regimes in both models are comparable. In
this regime $\mu_V/\mu_F$ is found to be well below~1, so that a
measurement of a value near~1 excludes this scenario. Moreover, a future
measurement of $\mu_{VV}$ with a precision of 5\% around the SM value
and the observation of $\mu_{\gamma\gamma} \gsim 0.9$ eliminates the
wrong-sign regime. These findings are consistent with the
observations in the 2HDM. \s

With the analysis tools provided in this study, the next natural step
would be to compare the N2HDM with other extended Higgs sectors
with a similar theoretical structure and Higgs spectrum, to investigate
which observables allow to distinguish (or to exclude) the models.

\subsubsection*{Acknowledgments}
The authors acknowledge financial support from the DAAD
project ``PPP Portugal 2015'' (ID: 57128671). M.S. was funded by FCT
through the grant SFRH/BPD/69971/2010. The work in this paper was also
supported by the CIDMA project UID/MAT/04106/2013.
We are grateful to Michael Spira for useful comments. We acknowledge
discussions with Pedro Ferreira.
\newpage
\section*{Appendix}
\setcounter{equation}{0}
\begin{appendix}
\section{N2HDM Parameters and Self-Couplings}
In the following we give the transformation formulae
between the N2HDM basis of the Lagrangian parameters and the
physical basis defined in Eq.~(\ref{eq:n2hdminputpars}). In
\ref{app:trilcoup} we give the trilinear self-couplings in terms of
the parameters of the physical basis.

\subsection{Basis Change \label{app:lamrels}}
With the definition of
\beq
\tilde{\mu}^2 = \frac{m_{12}^2}{s_\beta c_\beta}
\eeq
the quartic couplings $\lambda_i$ of the N2HDM potential can be
written in terms of the parameters of the physical basis as
\beq
\lambda_1 &=& \frac{1}{v^2 c_\beta^2} \left( -\tilde{\mu}^2 s_\beta^2
  + \sum_{i=1}^3 m_{H_i}^2 R_{i1}^2 \right) \label{eq:la1} \\
\lambda_2 &=& \frac{1}{v^2 s_\beta^2} \left( -\tilde{\mu}^2 c_\beta^2
 + \sum_{i=1}^3 m_{H_i}^2 R_{i2}^2  \right) \label{eq:la2}\\
\lambda_3 &=& \frac{1}{v^2} \left( -\tilde{\mu}^2
 + \frac{1}{s_\beta c_\beta}\sum_{i=1}^3 m_{H_i}^2 R_{i1} R_{i2} +
 2m_{H^\pm}^2 \right) \label{eq:la3}\\
\lambda_4 &=& \frac{1}{v^2} \left( \tilde{\mu}^2 + m_A^2 -
  2m_{H^\pm}^2 \right) \label{eq:la4}\\
\lambda_5 &=& \frac{1}{v^2} \left( \tilde{\mu}^2 - m_A^2 \right) \label{eq:la5}\\
\lambda_6 &=& \frac{1}{v_S^2} \sum_{i=1}^3 m_{H_i}^2 R_{i3}^2 \label{eq:la6}\\
\lambda_7 &=& \frac{1}{v v_S c_\beta} \sum_{i=1}^3 m_{H_i}^2 R_{i1}
R_{i3} \label{eq:la7}\\
\lambda_8 &=& \frac{1}{v v_S s_\beta} \sum_{i=1}^3 m_{H_i}^2 R_{i2}
R_{i3} \;. \label{eq:la8}
\eeq
The matrix elements $R_{ij}$ in terms of the mixing angles $\alpha_{1,2,3}$
have been defined in Eq.~(\ref{eq:mixingmatrix}).

\subsection{The Trilinear Higgs Self-Couplings \label{app:trilcoup}}
The trilinear Higgs self-couplings are derived from the terms of the
Higgs potential cubic in the Higgs fields. After rotation to the mass
basis and using the relations Eqs.~(\ref{eq:la1}) to (\ref{eq:la8})
the Feynman rules for the trilinear Higgs self-couplings between the
physical Higgs fields modulo a factor $i$ can be cast into the form
\beq
\hspace*{-0.7cm} g_{H_i AA} &=& \frac{1}{v} \left( - \tilde{\mu}^2 \left[
   \frac{R_{i1}}{c_\beta} + \frac{R_{i2}}{s_\beta} \right] +
 m_{H_i}^2 \left[ \frac{R_{i1} s_\beta^2}{c_\beta} + \frac{R_{i2}
     c_\beta^2}{s_\beta} \right] + 2m_A^2 [R_{i1} c_\beta + R_{i2}
 s_\beta] \right) \label{eq:self1} \\
\hspace*{-0.7cm} g_{H_i H^\pm H^\mp} &=& \frac{1}{v} \left( - \tilde{\mu}^2 \left[
   \frac{R_{i1}}{c_\beta} + \frac{R_{i2}}{s_\beta} \right] +
 m_{H_i}^2 \left[ \frac{R_{i1} s_\beta^2}{c_\beta} + \frac{R_{i2}
     c_\beta^2}{s_\beta} \right] + 2m_{H^\pm}^2 [R_{i1} c_\beta + R_{i2}
 s_\beta] \right) \label{eq:self2} \\
\hspace*{-0.7cm} g_{H_i H_i H_i} &=& \frac{3}{v} \left(-\tilde{\mu}^2 \left[ R_{i2}^2
    c_\beta \left( \frac{R_{i2} c_\beta}{s_\beta} - R_{i1} \right) +
    R_{i1}^2 s_\beta \left( \frac{R_{i1} s_\beta}{c_\beta} - R_{i2}
    \right) \right] \right.\nonumber \\
&+& \left.\frac{m_{H_i}^2}{v_S} \left[ R_{i3}^3 v + R_{i2}^3
    \frac{v_S}{s_\beta} + R_{i1}^3 \frac{v_S}{c_\beta} \right]
\right) \label{eq:self3} \\
\hspace*{-0.7cm} g_{H_i H_i H_j} &=& \frac{1}{v} \left( -\frac{1}{2}
  \tilde{\mu}^2 \left( \frac{R_{i2}}{s_\beta} -
      \frac{R_{i1}}{c_\beta} \right) \left( 6 R_{i2} R_{j2} + 6 R_{i3}
    R_{j3} s_\beta^2 + \sum_k \epsilon_{ijk} R_{k3} s_{2\beta} \right)
\right. \nonumber \\
&+& \left. \frac{2m_{H_i}^2 + m_{H_j}^2}{v_S} \left[ R_{i3}^2 R_{j3} v +
  R_{i2}^2 R_{j2} \frac{v_S}{s_\beta} + R_{i1}^2 R_{j1}
  \frac{v_S}{c_\beta}  \right] \right) \label{eq:self4} \\
\hspace*{-0.7cm} g_{H_1 H_2 H_3} &=& \frac{1}{v} \left( \tilde{\mu}^2
  \left[ (2 R_{12} R_{13} + R_{32} R_{33} ) c_\beta + (R_{31} R_{33} -
    3 R_{12} R_{23} R_{33} - R_{21} R_{23}) s_\beta
  \right. \right. \nonumber \\
&& + \left. 3 R_{12} R_{22} \left( \frac{R_{31}}{c_\beta} -
    \frac{R_{32}}{s_\beta} \right) + 3 R_{13} R_{23} R_{31}
  \frac{s_\beta^2}{c_\beta} \right] \nonumber \\
&+& \left. \frac{\sum_{i=1}^3 m_{H_i}^2 }{v_S} \left[ R_{13} R_{23}
    R_{33} v + R_{12} R_{22} R_{32} \frac{v_S}{s_\beta} - R_{11}
    (R_{22} R_{32} + R_{23} R_{33} ) \frac{v_S}{c_\beta} \right]
\right) \label{eq:self5}
\eeq
where $\epsilon_{ijk}$ denotes the totally antisymmetric tensor with
$\epsilon_{123}=1$. Note, that in Eqs.~(\ref{eq:self1}) to (\ref{eq:self5}) there is no
summation over repeated indices. The sums of different powers of
$R_{ij}$ arise from simplifications that exploit the orthogonality of
the mixing matrix. The employed formula reads
\beq
R_{ij} = (-1)^{i+j} \mbox{det} (\slash{\!\!\!\!R}_{ij}) \;.
\eeq
The matrix $\slash{\!\!\!\!R}_{ij}$ is the submatrix formed by deleting the
$i$-th row and the $j$-th column from $R$. The indices $i$ and $j$
take any values in $\{ 1,2,3 \}$.

\section{The Fortran Code {\tt N2HDECAY}  \label{app:n2hdecay}}
The code {\tt N2HDECAY} is the N2HDM implementation in the program
{\tt HDECAY}, written in Fortran77. It is based  on {\tt HDECAY
  v6.51}. The code is completely self-contained. All changes related
to the N2HDM have been implemented in the main file
{\tt n2hdecay.f}. Further linked routines have been taken over from the
original {\tt HDECAY} code. The implemented decay widths include the
most important state-of-the-art higher order QCD corrections and the
important off-shell decays. They can be taken over from the SM and the
minimal supersymmetric extension (MSSM), respectively, for which {\tt
  HDECAY} was originally designed. The electroweak corrections have been
consistently turned off as they cannot be taken over from the
available corrections in the SM and/or MSSM. \s

The N2HDM input parameters are specified in the input file {\tt
  n2hdecay.in}. It is based on the extension of the input file {\tt
  hdecay.in}. By setting the input value {\tt N2HDM}$=1$ in {\tt
  n2hdecay.in} the user chooses to calculate the N2HDM branching
ratios and total widths. The N2HDM-specific input parameters in the
physical basis are then set in the blocks `{\tt 2 Higgs Doublet Model}'
and `{\tt N2HDM}'. In the first block the user has to set {\tt PARAM=1}
and can choose the {\tt TYPE} of the fermion sector
symmetry. Furthermore, the parameters $\tan\beta$ and $m_{12}^2$ and
the pseudoscalar and charged Higgs masses are set here. In the {\tt
  N2HDM} block the neutral Higgs masses, the mixing angles and the VEV
of the singlet are set. We display part of an example input file
relevant for the N2HDM. The gray lines contain parameters not used in
the N2HDM.

\vspace*{0.2cm}
\begin{Verbatim}[fontsize=\small,commandchars=\\\{\},xleftmargin=10mm]
N2HDM    = 1
...
************************** 2 Higgs Doublet Model *************************
  TYPE: 1 (I), 2 (II), 3 (lepton-specific), 4 (flipped)
  PARAM: 1 (masses), 2 (lambda\_i)

PARAM    = 1
TYPE     = 2
********************
TGBET2HDM= 1.17639226D0
M\_12^2   = 3.28390121D5
******************** PARAM=1:
\textcolor{gray}{ALPHA\_H  = 10.D0}
\textcolor{gray}{MHL      = 10.D0}
\textcolor{gray}{MHH      = 10.D0}
MHA      = 9.02919728D2
MH+-     = 8.59398112D2
******************** PARAM=2:
\textcolor{gray}{LAMBDA1  = 0D0}
\textcolor{gray}{LAMBDA2  = 0D0}
\textcolor{gray}{LAMBDA3  = 0D0}
\textcolor{gray}{LAMBDA4  = 0D0}
\textcolor{gray}{LAMBDA5  = 0D0}
**************************** N2HDM ***********************************
*** needs TYPE, TGBET2HDM, M12\^2, MHA and MH+- from the 2HDM block ***
MH1      = 1.25090000D2
MH2      = 8.17422761D2
MH3      = 9.76339405D2
alpha1  = 0.79503834
alpha2  = 0.13549279
alpha3  = 1.46729273
V\_SING   = 1.49629673D3
**************************************************************************
...
\end{Verbatim}

The code is compiled with the {\tt makefile} by typing {\tt
  make}. This produces an executable file called {\tt run}. Typing
{\tt run} executes the program, which calculates the branching ratios
and total widths that are written out together with the mass of the
decaying Higgs boson. The output files are called {\tt
  br.X\_N2HDM\_y}. Here {\tt X=H1, H2, H3, A, H+} denotes the decaying
Higgs particle. Files with the suffix {\tt y=a} contain the branching
ratios into fermions, with {\tt y=b} the ones into gauge bosons and
the ones with {\tt y=c, d} the branching ratios into lighter Higgs
pairs or a Higgs-gauge boson final state. In the following we present
the example of an output file as obtained from the above input file. The
produced output in the four output files {\tt br.H3\_N2HDM\_y} for the
heaviest neutral Higgs boson is given by

\vspace*{0.2cm}
\begin{verbatim}
   MH3         BB       TAU TAU     MU MU         SS         CC         TT
---------------------------------------------------------------------------
 976.339     0.3458E-03 0.5450E-04 0.1927E-06 0.1259E-06 0.1267E-04 0.8026

   MH3           GG     GAM GAM     Z GAM         WW         ZZ
---------------------------------------------------------------------------
 976.339     0.1326E-02 0.3417E-05 0.6717E-06 0.4762E-01 0.2350E-01

   MH3          H1H1       H1H2       H2H2        AA        Z A
---------------------------------------------------------------------------
 976.339     0.6375E-01 0.2923E-03  0.000     0.9115E-13 0.7821E-04

   MH3       W+- H-+     H+ H-      WIDTH
---------------------------------------------------------------------------
 976.339     0.6038E-01  0.000      43.40
\end{verbatim}

All files necessary for the program can be downloaded at the url: \\
\centerline{\tt http://www.itp.kit.edu/$\sim$maggie/N2HDECAY} \\
The webpage contains a short explanation of the program and
information on updates and modifications of the program. Furthermore,
sample output files can be found for a given input.

\section{Global Minimum Conditions \label{sec:appglobal}}
We start by recalling that, up to gauge symmetries, the most general constant field configuration, where all fields are real, is
\beq
\langle \Phi_1 \rangle = \left( \begin{array}{c} 0 \\ v_1 \end{array}
\right) \;, \quad
\langle \Phi_2 \rangle = \left( \begin{array}{c} v_{\text{cb}} \\ v_2
    + i v_{\text{cp}} \end{array} \right) \;, \quad
\langle \Phi_S \rangle = v_S \; . \label{eq:appfieldconfig}
\eeq
The subscripts cp and cb refer to the case where CP or charge are
spontaneously broken in scenarios where also both $v_1$ and $v_2$ are
non-zero. In the case where only $v_{\text{cp}}$ and/or $v_{\text{cb}}$
are non-zero there is in fact no CP or charge breaking. One way to see this is by noting that such a configuration is continuously connected by a gauge transformation to the case where only $v_2\neq 0$ (where it is more clear that there is no CP or charge breaking). Such a gauge transformation amounts to a redefinition of the charge operator by a rotation. \s

In order to find all possible minima we consider the stationarity
conditions for the VEVs,
\begin{align}
\left\langle \frac{\partial V}{\partial v_1} \right\rangle
&= 0 \Leftrightarrow & v_2 m_{12}^2
- v_1 m_{11}^2 &= \frac{1}{2} v_1 (
v_1^2 \lambda_1 + v_2^2 \lambda_{345} + v_{\text{cb}}^2 \lambda_3 +
v_{\text{cp}}^2 \lambda_{34-5} + v_S^2 \lambda_7 )
\label{eq:v1cond} \\
\left\langle \frac{\partial V}{\partial v_2} \right\rangle
&= 0 \Leftrightarrow & v_1 m_{12}^2
- v_2 m_{22}^2 &= \frac{1}{2} v_2 (
v_1^2 \lambda_{345} + v_2^2 \lambda_2 + v_{\text{cb}}^2 \lambda_2 +
v_{\text{cp}}^2 \lambda_{2} + v_S^2 \lambda_8 ) \label{eq:v2cond} \\
\left\langle \frac{\partial V}{\partial v_{\text{cb}}} \right\rangle
&= 0 \Leftrightarrow & -
v_{\text{cb}} m_{22}^2 &= \frac{1}{2}
v_{\text{cb}} ( v_1^2 \lambda_{3} + v_2^2 \lambda_2 + v_{\text{cb}}^2 \lambda_2 +
v_{\text{cp}}^2 \lambda_{2} + v_S^2 \lambda_8 ) \label{eq:vcbcond} \\
\left\langle \frac{\partial V}{\partial v_{\text{cp}}} \right\rangle
&= 0 \Leftrightarrow & -
v_{\text{cp}} m_{22}^2 &= \frac{1}{2}
v_{\text{cp}} ( v_1^2 \lambda_{34-5} + v_2^2 \lambda_2 +
v_{\text{cb}}^2 \lambda_2 + v_{\text{cp}}^2 \lambda_{2} + v_S^2
\lambda_8 ) \label{eq:vcpcond}
\end{align}
\begin{align}
\left\langle \frac{\partial V}{\partial v_{\text{S}}} \right\rangle
&= 0 \Leftrightarrow & -
v_{\text{S}} m_{S}^2 &= \frac{1}{2}
v_{\text{S}} ( v_1^2 \lambda_{7} + v_2^2 \lambda_8 +
v_{\text{cb}}^2 \lambda_8 + v_{\text{cp}}^2 \lambda_{8} + v_S^2
\lambda_6 ) \;, \label{eq:vscond}
\end{align}
where $\lambda_{345}$ has been defined in Eq.~(\ref{eq:l345}) and
\beq
\lambda_{34-5} \equiv \lambda_3 + \lambda_4 - \lambda_5 \;.
\eeq
The derivatives with respect to the degrees of freedom, which have been removed through a gauge transformation in Eq.~(\ref{eq:appfieldconfig}), contribute with three further conditions
\begin{align}
  0 &= v_{\text{cb}}  \left(v_1 v_2(\lambda_4 + \lambda_5) - 2 m_{12}^2\right)\label{eq:vcbRPhi1}\\
0 &= v_1 v_\text{cb} v_\text{cp} \left(\lambda_4 -\lambda_5\right)\label{eq:vcbIPhi1}\\
0 &=  v_\text{cp}\left( v_1 v_2 \lambda_5 - m_{12}^2\right)\; .\label{eq:vcpPhi1}
\end{align}
From Eqs.~(\ref{eq:vcbcond}) and (\ref{eq:vcpcond}) we infer that
except for the special case
\beq
\lambda_4 = \lambda_5
\eeq
the VEVs $v_{\text{cb}}$ and $v_{\text{cp}}$ cannot be simultaneously
non-zero. Eq.~(\ref{eq:vcbIPhi1}) is therefore always trivially satisfied. From Eqs.~(\ref{eq:v1cond}) and (\ref{eq:v2cond}) we conclude that
\beq
(v_1 = 0 \; \Leftrightarrow \; v_2=0) \; \vee \; m_{12}^2 = 0 \;.
\eeq
We further observe from Eqs.~(\ref{eq:vcbRPhi1}) and (\ref{eq:vcpPhi1}) that
\begin{align}
  v_1=v_2=0 \;\Rightarrow\; \left(v_\text{cb}=v_\text{cp}=0 \; \vee\; m_{12}^2 = 0\right)\;.
\end{align}
The configurations forbidden by non-vanishing $m_{12}^2$ are inert
stationary points where the whole doublet VEV can be brought into one
Higgs doublet through a basis transformation.
Models with inert Higgs doublets show a very different phenomenology
and we only consider points with
\beq
|m_{12}^2| > 0\;.
\eeq
In this case $v_1$ and $v_2$ are either both non-zero or $v_S$ is the only non-vanishing VEV.
All corresponding possible combinations of VEVs, {\it i.e.}~all possible
stationary points, are listed in table~\ref{tab:vevcombo}.
The global minimum is given by the stationary point that leads to the
smallest value for the scalar potential $V$.
Since we have defined the field configurations
Eq.~(\ref{eq:appfieldconfig}) such that
the VEVs $v_1$, $v_2$, $v_{\text{cp}}$ and $v_{\text{cb}}$ are real,
only solutions of the stationarity conditions with all
VEVs squared positive are allowed. In the following we will provide the
values of the scalar potential at all the stationary points listed in
table~\ref{tab:vevcombo}.
These stationary points need only be considered if the corresponding
positivity conditions on the VEVs squared are satisfied.
\begin{table}
\begin{center}
\begin{tabular}{cccccccc}
\toprule
Case            & I & IIa & IIb & sI & sIIa & sIIb & s \\
\midrule
$v_1$           & 1 & 1 & 1 & 1 & 1 & 1 & 0 \\
$v_2$           & 1 & 1 & 1 & 1 & 1 & 1 & 0 \\
$v_{\text{cp}}$ & 0 & 1 & 0 & 0 & 1 & 0 & 0 \\
$v_{\text{cb}}$ & 0 & 0 & 1 & 0 & 0 & 1 & 0 \\
$v_S$           & 0 & 0 & 0 & 1 & 1 & 1 & 1 \\
\bottomrule
\end{tabular}
  \caption{All possible cases of VEVs being zero (0) or non-zero
   (1). Cases that are allowed only when specific parameter
   configurations hold, are not shown. \label{tab:vevcombo}}
\end{center}
\end{table}

\subsection{2HDM-like stationary points}
The three cases of the 2HDM-like stationary points are obtained setting
$v_S=0$. These are the stationary points of a 2HDM potential with the
same parameters $m_{11}^2$, $m_{22}^2$, $m_{12}^2$ and
$\lambda_{1-5}$.  \s

\vspace*{0.2cm}
\noindent {\bf Case I:} \s

\noindent
This case with a CP and charge conserving minimum
is the most complicated one. We start by rewriting the minimum
conditions in terms of
\beq
v_1 = v \cos \delta \quad \mbox{and} \quad
v_2 = v \sin \delta \;. \label{eq:v12vals}
\eeq
Without loss of generality all VEVs except $v_2$ can be chosen
positive due to  the $\mathbb{Z}_2$ symmetry. The convention $v_1 >0$ implies
\beq
-\frac{\pi}{2} < \delta < \frac{\pi}{2} \;. \label{eq:sindelinterv}
\eeq
The resulting system of equations is used to eliminate $v$ leading to
a single quartic equation for $\sin^2 \delta$,
\beq
0 &=& (m_{12}^2)^2 \lambda_1^2 \nonumber \\
&+& \sin^2 \delta \big[-(m_{11}^2 \lambda_{345} - m_{22}^2 \lambda_1)^2 -
4 (m_{12}^2)^2 \lambda_1^2\big] \nonumber \\
&+& (\sin^2 \delta)^2 \big[3(m_{11}^2 \lambda_{345} - m_{22}^2
\lambda_1)^2 + 2(m_{11}^2\lambda_{345}-m_{22}^2 \lambda_1)(m_{22}^2
\lambda_{345} -m_{11}^2 \lambda_2) \nonumber \\
&& +2(m_{12}^2)^2 (3\lambda_1^2 - \lambda_1 \lambda_2)\big] \nonumber \\
&+& (\sin^2 \delta)^3 \big[-3 (m_{11}^2 \lambda_{345} - m_{22}^2
\lambda_1)^2 + 4 (m_{12}^2)^2 \lambda_1 (\lambda_2 - \lambda_1)
\nonumber \\
&& - (m_{22}^2 \lambda_{345} - m_{11}^2 \lambda_2) ((4m_{11}^2 +
m_{22}^2)\lambda_{345} - 4m_{22}^2 \lambda_1 - m_{11}^2 \lambda_2) \big]
\nonumber \\
&+& (\sin^2 \delta)^4 \big[(m_{22}^2(\lambda_1-\lambda_{345}) + m_{11}^2 (\lambda_2 - \lambda_{345}))^2 \nonumber \\
&&+(m_{12}^2)^2 (\lambda_1 - \lambda_2)^2\big]\;.
\label{eq:sindeltasol}
\eeq
Among the four solutions of Eq.~(\ref{eq:sindeltasol}) for $\sin^2
\delta$, which can be obtained numerically, only those are allowed
that are real and in the open interval (0,1). Reality for $\delta$ is
required so that $v_1$ and $v_2$ are real. Each of these solutions leads to two
possible values for $v^2$,
\beq
v^2 = \frac{2 \left(m_{22}^2\mp m_{12}^2 \sqrt{\frac{1}{\sin^2
        \delta}-1} \right)}{\lambda_{345}(\sin^2 \delta -1) -
  \lambda_2 \sin^2 \delta} \;.
\eeq
The $\mp$ signs correspond to the two possible signs of $\sin
\delta$ in the interval Eq.~(\ref{eq:sindelinterv}). The values for
$v_1$ and $v_2$ are then obtained from Eq.~(\ref{eq:v12vals}) where
both possible signs of $v_2$ need to be considered. Altogether this
yields up to 16 possible solutions for $v_1$ and $v_2$ given by the
four solutions for $\sin^2 \delta$ times two for the sign of $\sin
\delta$ times two for the sign of $v_2$. It should be noted only two of these solutions are independent as shown in~\cite{Barroso:2007rr, Ivanov:2007de}.\s

\vspace*{0.2cm}
\noindent
{\bf Case II:} \s

\noindent
The system of minimum conditions obtained here can be solved
analytically both for case IIa and IIb. The formula for the value
of the potential at these points can be cast into the form
\beq
V(\mbox{II}) &=& \frac{(m_{11}^2)^2 \lambda_2 - 2 m_{11}^2 m_{22}^2x +
  (m_{22}^2)^2 \lambda_1}{2 \Lambda_{12}^{xx}} -
\frac{(m_{12}^2)^2}{\lambda_{345}-x} \;,
\eeq
where
\beq
x = \left\{ \begin{array}{ll} \lambda_{34-5} & \mbox{ in case IIa} \\
\lambda_3 & \mbox{ in case IIb}
\end{array} \right.
\eeq
and
\beq
\Lambda_{kl}^{ij} \equiv \lambda_i \lambda_j - \lambda_k \lambda_l \;.
\eeq
Simultaneously the positivity conditions for the squared VEVs have to be
fulfilled. They read
\beq
0 < v_{\text{cb}}^2 \text{ or } v_{\text{cp}}^2 &=& \frac{2(m_{12}^2)^2
  \Lambda^{xx}_{12}}{(\lambda_{345}-x)^2 (m_{22}^2 x-m_{11}^2
  \lambda_2)} + \frac{2(m_{22}^2\lambda_1-
  m_{11}^2x)}{\Lambda_{12}^{xx}} \\
0 < v_1^2 &=&  \frac{2(m_{11}^2 \lambda_2- m_{22}^2
  x)}{\Lambda_{12}^{xx}}  \;.
\eeq
Positivity of $v_2^2$ is guaranteed if $v_1^2$ is positive. \s

\vspace*{0.2cm}
\noindent {\bf Special case $\lambda_4 = \lambda_5$, $v_S=0$:} \s

\noindent
In this special case $v_\text{cb}$ and $v_\text{cp}$ can be simultaneously non-zero.
This is only possible if $v_1$ and $v_2$ are also non-zero. The resulting stationary value is
\beq
V(\lambda_4 = \lambda_5, v_S=0) =-\frac{(m_{12}^2)^2}{2\lambda_4}+\frac{m_{11}^2\lambda_2 +m_{22}^2\lambda_1 - 2m_{11}^2m_{22}^2\lambda_3}{\Lambda_{12}^{33}}\;.
\eeq
This value is obtained on a ring of constant $v_\text{cp}^2+v_\text{cb}^2$. The positivity conditions therefore read
\beq
0 < v_{\text{cb}}^2 + v_{\text{cp}}^2 &=&  -\frac{(m_{12}^2)^2\Lambda^{33}_{12}}{2(m_{11}^2\lambda_2-m_{22}^2\lambda_3)\lambda_4^2}-\frac{2 (m_{11}^2\lambda_3-m_{22}^2\lambda_1)}{\Lambda_{12}^{33}}\\
0 < v_1^2 &=& \frac{2(m_{11}^2\lambda_2-m_{22}^2\lambda_3)}{\Lambda_{12}^{33}}\;.
\eeq

\subsection{Stationary points with a singlet VEV}
The stationary points for a non-vanishing singlet VEV $v_S \ne 0$, as
possible in the N2HDM, are obtained analogously to the previous cases. \s

\vspace*{0.2cm}
\noindent {\bf Case sI:} \s

\noindent
Again the case with non-vanishing $v_1$ and $v_2$ but zero
$v_{\text{cp}}$ and $v_{\text{cb}}$ is the most complicated one and
leads to a quartic equation for $\sin^2 \delta$, with $\delta$ defined
in Eq.~(\ref{eq:v12vals}). It reads
\beq
0 &=& (m_{12}^2)^2 (\Lambda_{16}^{77})^2 \nonumber \\
&+& \sin^2 \delta \big[-\big((m_{11}^2 \lambda_6 - m_S^2 \lambda_7)
\lambda_{345} + m_{22}^2 \Lambda_{16}^{77} + (m_S^2 \lambda_1 -
m_{11}^2 \lambda_7) \lambda_8\big)^2-4(m_{12}^2)^2 (\Lambda_{16}^{77})^2\big]
\nonumber \\
&+& (\sin^2 \delta)^2 \big[\big((m_{11}^2 \lambda_6 - m_S^2
\lambda_7)\lambda_{345} -m_{11}^2 \lambda_7 \lambda_8 + m_{22}^2
\Lambda_{16}^{77} + m_S^2 \lambda_1 \lambda_8 \big) \nonumber \\
&& \times \big((3m_{11}^2 \lambda_6 + 2 m_{22}^2 \lambda_6 -m_S^2 (3
\lambda_7 + 2 \lambda_8))\lambda_{345} + m_{11}^2 (2 \Lambda_{26}^{88}
- 3 \lambda_7 \lambda_8) \nonumber\\
&&  + m_{22}^2 (3\Lambda_{16}^{77} - 2 \lambda_7
\lambda_8) + m_S^2 (2 \lambda_2 \lambda_7 + 3 \lambda_1 \lambda_8)\big)
+ 2 (m_{12}^2)^2 \Lambda_{16}^{77} (3 \Lambda_{16}^{77}-
\Lambda_{26}^{88})\big] \nonumber \\
&+& (\sin^2 \delta)^3 \big[\big((m_S^2 (3\lambda_7 + \lambda_8)-(3m_{11}^2 +
m_{22}^2) \lambda_6)\lambda_{345} + m_{11}^2 (3 \lambda_7 \lambda_8 -
\Lambda_{26}^{88} ) \nonumber \\
&& + m_{22}^2 (\lambda_7 \lambda_8 - 3 \Lambda_{16}^{77}) - m_S^2
(\lambda_2 \lambda_7 + 3 \lambda_1 \lambda_8)\big) \nonumber \\
&& \times \big(((m_{11}^2+m_{22}^2 )\lambda_6 - m_S^2 (\lambda_7 +
\lambda_8))\lambda_{345} + m_{11}^2
(\Lambda_{26}^{88}-\lambda_7\lambda_8) \nonumber \\
&& + m_{22}^2 (\Lambda_{16}^{77} - \lambda_7 \lambda_8) + m_S^2
(\lambda_2 \lambda_7 + \lambda_1 \lambda_8)\big) \nonumber + 4
(m_{12}^2)^2 \Lambda_{16}^{77}
(\Lambda_{26}^{88}-\Lambda_{16}^{77} )\big] \nonumber
\eeq
\beq
&+& (\sin^2 \delta)^4 \big[\big(((m_{11}^2 + m_{22}^2)\lambda_6 -m_S^2
(\lambda_7+ \lambda_8))\lambda_{345} + m_{11}^2 (\Lambda_{26}^{88} -
\lambda_7 \lambda_8) \nonumber \\
&& + m_{22}^2 (\Lambda_{16}^{77} - \lambda_7 \lambda_8) + m_S^2
(\lambda_2 \lambda_7 + \lambda_1 \lambda_8)\big)^2 + (m_{12}^2)^2
(\Lambda_{26}^{88}-\Lambda_{16}^{77})^2 \big] \;.
\eeq
The real solutions for $\sin^2\delta$ in the open interval $(0,1)$
imply two possible values for $v^2$,
\beq
v^2 = \frac{2 \lambda_6 \left( m_{22}^2 \mp m_{12}^2
    \sqrt{\frac{1}{\sin^2 \delta}-1}\right) - 2m_S^2
  \lambda_8}{(\sin^2 \delta -1)(\lambda_6 \lambda_{345}-\lambda_7
  \lambda_8) + \Lambda_{26}^{88} \sin^2 \delta} \;,
\eeq
depending on the sign of $\sin\delta$ in the interval. For $v_S^2$ we
have
\beq
v_S^2 = -\frac{\lambda_7 v_1^2 + \lambda_8 v_2^2 +2m_S^2}{\lambda_6} \;.
\eeq
Valid solutions are given for positive values of $v^2$ and
$v_S^2$. The sign of $v_S$ is irrelevant so that also here we have up
to 16 solutions and the corresponding values of the potential that can be
obtained numerically. \s

\vspace*{0.2cm}
\noindent {\bf Case sII:} \s

\noindent
Defining again for the two subcases
\beq
x = \left\{ \begin{array}{ll} \lambda_{34-5} & \mbox{ in case sIIa} \\
\lambda_3 & \mbox{ in case sIIb} \end{array} \right. \;,
\eeq
we have the stationary values and positivity conditions for the case sII stationary point given by
\beq
V(\mbox{sII}) &=& -\frac{(m_{12}^2)^2}{\lambda_{345}-x} +
\frac{(m_{11}^2)^2 \Lambda_{26}^{88}+(m_{22}^2)^2 \Lambda_{16}^{77} +
  (m_S^2)^2 \Lambda_{12}^{xx}}{2(\lambda_7 \Lambda^{x8}_{27}+x
  \Lambda^{78}_{x6}- \lambda_1 \Lambda_{26}^{88})} \nonumber \\
&& - \frac{m_{11}^2 m_{22}^2 \Lambda_{x6}^{78}+ m_{11}^2 m_S^2
  \Lambda_{27}^{x8} + m_{22}^2 m_S^2 \Lambda_{18}^{x7}}{\lambda_7
  \Lambda_{27}^{x8}+x \Lambda_{x6}^{78} - \lambda_1 \Lambda_{26}^{88}}
\eeq
and
\beq
0 < v_{\text{cb}}^2 \text{ or } v_{\text{cp}}^2 &=& \frac{(m_{12}^2)^2
  (\lambda_7 \Lambda_{27}^{x8} + x \Lambda_{x6}^{78} - \lambda_1
  \Lambda_{26}^{88})}{2(\lambda_{345}-x)^2 (m_S^2 \Lambda_{27}^{x8} +
  m_{22}^2 \Lambda_{x6}^{78} - m_{11}^2 \Lambda_{26}^{88})} \nonumber
\\
&& + \frac{2m_{22}^2 \Lambda_{16}^{77} -2 m_S^2 \Lambda_{18}^{x7} -
  2m_{11}^2 \Lambda_{x6}^{78}}{\lambda_7 \Lambda_{27}^{x8} + x
  \Lambda_{x6}^{78} - \lambda_1 \Lambda_{26}^{88}} \\
0 < v_S^2 &=& \frac{2m_S^2 \Lambda_{12}^{xx}-2m_{11}^2
  \Lambda_{27}^{x8}- 2m_{22}^2 \Lambda_{18}^{x7}}{\lambda_7
  \Lambda_{27}^{x8} + x \Lambda_{x6}^{78} - \lambda_1
  \Lambda_{26}^{88}} \\
0 < v_1^2 &=& \frac{-2 m_S^2 \Lambda_{27}^{x8}-2m_{22}^2
  \Lambda_{x6}^{78} + 2 m_{11}^2 \Lambda_{26}^{88}}{\lambda_7
  \Lambda_{27}^{x8} + x \Lambda_{x6}^{78} - \lambda_1
  \Lambda_{26}^{88}} \;.
\eeq
Positivity of $v_2^2$ is ensured if $v_1^2$ is positive. \s

\vspace*{0.2cm}
\noindent {\bf Case s:} \s

\noindent
For the case denoted by s the value of the potential at the stationary
point simplifies to
\beq
V(\mbox{sIV}) = - \frac{(m_S^2)^2}{2\lambda_6} \;,
\eeq
which is a valid solution ($v_S^2>0$) if
\beq
m_S^2 < 0 \;.
\eeq

\vspace*{0.2cm}
\noindent {\bf Special case $\lambda_4 = \lambda_5$, $v_S\neq0$:} \s

\noindent
If $\lambda_4 = \lambda_5$ all VEVs can be simultaneously non-zero. In
this case the scalar potential takes the stationary value
\beq
V(\lambda_4 = \lambda_5, v_S\neq0) &=& - \frac{(m_{12}^2)^2}{2 \lambda_4} +
\frac{(m_{11}^2)^2 \Lambda_{26}^{88} +
  (m_{22}^2)^2 \Lambda_{16}^{77} + (m_S^2)^2 \Lambda_{12}^{33} }{2 (\lambda_7 \Lambda_{27}^{38}+
  \lambda_3 \Lambda_{36}^{78} - \lambda_1 \Lambda_{26}^{88})}\nonumber\\
  &&-\frac{m_{11}^2m_{22}^2\Lambda_{36}^{78}+m_{11}^2m_S^2\Lambda_{27}^{38}+m_{22}^2m_S^2\Lambda^{37}_{18}}{\lambda_7 \Lambda_{27}^{38}+
  \lambda_3 \Lambda_{36}^{78} - \lambda_1 \Lambda_{26}^{88}} \;.
\eeq
This stationary value is again obtained on a ring of constant $v_\text{cb}^2+v_\text{cp}^2$. The corresponding positivity conditions read
\beq
\hspace*{-0.5cm}0 < v_{\text{cb}}^2 + v_{\text{cp}}^2 &=& \frac{(m_{12}^2)^2 (
  \lambda_7 \Lambda_{27}^{38}+\lambda_3 \Lambda_{36}^{78}-\lambda_1
  \Lambda_{26}^{88})}{2\lambda_4^2 (m_S^2 \Lambda_{27}^{38} + m_{22}^2
  \Lambda_{36}^{78} -m_{11}^2 \Lambda_{26}^{88})} + \frac{2(m_{22}^2
  \Lambda_{16}^{77} -  m_S^2 \Lambda_{18}^{37}- m_{11}^2
  \Lambda_{36}^{78})}{\lambda_7 \Lambda_{27}^{38}+\lambda_3 \Lambda_{36}^{78} -\lambda_1
  \Lambda_{26}^{88}} \\
\hspace*{-0.5cm}0 < v_S^2 &=& \frac{2 (m_S^2 \Lambda_{12}^{33} - m_{11}^2
  \Lambda_{27}^{38} - m_{22}^2 \Lambda_{18}^{37})}{\lambda_7
  \Lambda_{27}^{38} + \lambda_3 \Lambda_{36}^{78} - \lambda_1
  \Lambda_{26}^{88}} \\
\hspace*{-0.5cm}0 < v_1^2 &=& -\frac{2(m_S^2 \Lambda_{27}^{38}+ m_{22}^2
  \Lambda_{36}^{78} - m_{11}^2 \Lambda_{26}^{88})}{\lambda_7
  \Lambda_{27}^{38} + \lambda_3 \Lambda_{36}^{78} - \lambda_1
  \Lambda_{26}^{88}} \;.
\eeq

In order to check if our minimum is the global one we compare the
value of the scalar potential at our minimum with the values of the
potential at all the stationary points listed above. In practice this
means that we need to compare with the up to five different
analytically known values of cases (s)II and case s and with the
numerical solutions of case (s)I.

\end{appendix}

\clearpage
\vspace*{1cm}
\bibliographystyle{h-physrev}

\begin{thebibliography}{100}

\bibitem{Aad:2012tfa}
ATLAS Collaboration, G.~Aad {\em et~al.},
\newblock Phys.Lett. {\bf B716}, 1 (2012), 1207.7214.

\bibitem{Chatrchyan:2012ufa}
CMS Collaboration, S.~Chatrchyan {\em et~al.},
\newblock Phys.Lett. {\bf B716}, 30 (2012), 1207.7235.

\bibitem{Englert:2014uua}
C.~Englert {\em et~al.},
\newblock J. Phys. {\bf G41}, 113001 (2014), 1403.7191.

\bibitem{Aad:2015zhl}
ATLAS, CMS, G.~Aad {\em et~al.},
\newblock Phys. Rev. Lett. {\bf 114}, 191803 (2015), 1503.07589.

\bibitem{Terazawa:1976xx}
H.~Terazawa, K.~Akama, and Y.~Chikashige,
\newblock Phys. Rev. {\bf D15}, 480 (1977).

\bibitem{Terazawa:1979pj}
H.~Terazawa,
\newblock Phys. Rev. {\bf D22}, 184 (1980).

\bibitem{Kaplan:1983fs}
D.~B. Kaplan and H.~Georgi,
\newblock Phys. Lett. {\bf B136}, 183 (1984).

\bibitem{Dimopoulos:1981xc}
S.~Dimopoulos and J.~Preskill,
\newblock Nucl. Phys. {\bf B199}, 206 (1982).

\bibitem{Banks:1984gj}
T.~Banks,
\newblock Nucl. Phys. {\bf B243}, 125 (1984).

\bibitem{Kaplan:1983sm}
D.~B. Kaplan, H.~Georgi, and S.~Dimopoulos,
\newblock Phys. Lett. {\bf B136}, 187 (1984).

\bibitem{Georgi:1984ef}
H.~Georgi, D.~B. Kaplan, and P.~Galison,
\newblock Phys. Lett. {\bf B143}, 152 (1984).

\bibitem{Georgi:1984af}
H.~Georgi and D.~B. Kaplan,
\newblock Phys. Lett. {\bf B145}, 216 (1984).

\bibitem{Dugan:1984hq}
M.~J. Dugan, H.~Georgi, and D.~B. Kaplan,
\newblock Nucl. Phys. {\bf B254}, 299 (1985).

\bibitem{Giudice:2007fh}
G.~F. Giudice, C.~Grojean, A.~Pomarol, and R.~Rattazzi,
\newblock JHEP {\bf 06}, 045 (2007), hep-ph/0703164.

\bibitem{Agashe:2004rs}
K.~Agashe, R.~Contino, and A.~Pomarol,
\newblock Nucl. Phys. {\bf B719}, 165 (2005), hep-ph/0412089.

\bibitem{Contino:2006qr}
R.~Contino, L.~Da~Rold, and A.~Pomarol,
\newblock Phys. Rev. {\bf D75}, 055014 (2007), hep-ph/0612048.

\bibitem{Khachatryan:2014kca}
CMS, V.~Khachatryan {\em et~al.},
\newblock Phys. Rev. {\bf D92}, 012004 (2015), 1411.3441.

\bibitem{Aad:2015mxa}
ATLAS, G.~Aad {\em et~al.},
\newblock Eur. Phys. J. {\bf C75}, 476 (2015), 1506.05669.

\bibitem{Khachatryan:2014jba}
CMS, V.~Khachatryan {\em et~al.},
\newblock Eur. Phys. J. {\bf C75}, 212 (2015), 1412.8662.

\bibitem{Aad:2015gba}
ATLAS, G.~Aad {\em et~al.},
\newblock Eur. Phys. J. {\bf C76}, 6 (2016), 1507.04548.

\bibitem{Gunion:1989we}
J.~F. Gunion, H.~E. Haber, G.~L. Kane, and S.~Dawson,
\newblock Front.Phys. {\bf 80}, 1 (2000).

\bibitem{Lee:1973iz}
T.~D. Lee,
\newblock Phys. Rev. {\bf D8}, 1226 (1973).

\bibitem{Branco:2011iw}
G.~C. Branco {\em et~al.},
\newblock Phys. Rept. {\bf 516}, 1 (2012), 1106.0034.

\bibitem{He:2008qm}
X.-G. He, T.~Li, X.-Q. Li, J.~Tandean, and H.-C. Tsai,
\newblock Phys. Rev. {\bf D79}, 023521 (2009), 0811.0658.

\bibitem{Grzadkowski:2009iz}
B.~Grzadkowski and P.~Osland,
\newblock Phys. Rev. {\bf D82}, 125026 (2010), 0910.4068.

\bibitem{Logan:2010nw}
H.~E. Logan,
\newblock Phys. Rev. {\bf D83}, 035022 (2011), 1010.4214.

\bibitem{Boucenna:2011hy}
M.~S. Boucenna and S.~Profumo,
\newblock Phys. Rev. {\bf D84}, 055011 (2011), 1106.3368.

\bibitem{He:2011gc}
X.-G. He, B.~Ren, and J.~Tandean,
\newblock Phys. Rev. {\bf D85}, 093019 (2012), 1112.6364.

\bibitem{Bai:2012nv}
Y.~Bai, V.~Barger, L.~L. Everett, and G.~Shaughnessy,
\newblock Phys. Rev. {\bf D88}, 015008 (2013), 1212.5604.

\bibitem{He:2013suk}
X.-G. He and J.~Tandean,
\newblock Phys. Rev. {\bf D88}, 013020 (2013), 1304.6058.

\bibitem{Cai:2013zga}
Y.~Cai and T.~Li,
\newblock Phys. Rev. {\bf D88}, 115004 (2013), 1308.5346.

\bibitem{Guo:2014bha}
J.~Guo and Z.~Kang,
\newblock Nucl.\ Phys.\ B {\bf 898} (2015) 415, arXiv:1401.5609.

\bibitem{Wang:2014elb}
L.~Wang and X.-F. Han,
\newblock Phys. Lett. {\bf B739}, 416 (2014), 1406.3598.

\bibitem{Drozd:2014yla}
A.~Drozd, B.~Grzadkowski, J.~F. Gunion, and Y.~Jiang,
\newblock JHEP {\bf 11}, 105 (2014), 1408.2106.

\bibitem{Campbell:2015fra}
R.~Campbell, S.~Godfrey, H.~E. Logan, A.~D. Peterson, and A.~Poulin,
\newblock Phys. Rev. {\bf D92}, 055031 (2015), 1505.01793.

\bibitem{Drozd:2015gda}
  A.~Drozd, B.~Grzadkowski, J.~F.~Gunion and Y.~Jiang,
\newblock JCAP {\bf 1610} (2016) no.10,  040, arXiv:1510.07053.

\bibitem{vonBuddenbrock:2016rmr}
S.~von Buddenbrock {\em et~al.},
\newblock (2016), 1606.01674.

\bibitem{Chen:2013jvg}
C.-Y. Chen, M.~Freid, and M.~Sher,
\newblock Phys. Rev. {\bf D89}, 075009 (2014), 1312.3949.

\bibitem{Djouadi:1999gv}
A.~Djouadi, W.~Kilian, M.~Muhlleitner, and P.~M. Zerwas,
\newblock Eur. Phys. J. {\bf C10}, 27 (1999), hep-ph/9903229.

\bibitem{Djouadi:1999rca}
A.~Djouadi, W.~Kilian, M.~Muhlleitner, and P.~M. Zerwas,
\newblock Eur. Phys. J. {\bf C10}, 45 (1999), hep-ph/9904287.

\bibitem{Muhlleitner:2000jj}
M.~M. Muhlleitner,
{\em {Higgs particles in the standard model and supersymmetric
  theories}},
\newblock PhD thesis, Hamburg U., 2000, hep-ph/0008127.

\bibitem{Fayet:1974pd}
P.~Fayet,
\newblock Nucl.Phys. {\bf B90}, 104 (1975).

\bibitem{Barbieri:1982eh}
R.~Barbieri, S.~Ferrara, and C.~A. Savoy,
\newblock Phys.Lett. {\bf B119}, 343 (1982).

\bibitem{Dine:1981rt}
M.~Dine, W.~Fischler, and M.~Srednicki,
\newblock Phys.Lett. {\bf B104}, 199 (1981).

\bibitem{Nilles:1982dy}
H.~P. Nilles, M.~Srednicki, and D.~Wyler,
\newblock Phys.Lett. {\bf B120}, 346 (1983).

\bibitem{Frere:1983ag}
J.~Frere, D.~Jones, and S.~Raby,
\newblock Nucl.Phys. {\bf B222}, 11 (1983).

\bibitem{Derendinger:1983bz}
J.~Derendinger and C.~A. Savoy,
\newblock Nucl.Phys. {\bf B237}, 307 (1984).

\bibitem{Ellis:1988er}
J.~R. Ellis, J.~Gunion, H.~E. Haber, L.~Roszkowski, and F.~Zwirner,
\newblock Phys.Rev. {\bf D39}, 844 (1989).

\bibitem{Drees:1988fc}
M.~Drees,
\newblock Int.J.Mod.Phys. {\bf A4}, 3635 (1989).

\bibitem{Ellwanger:1993xa}
U.~Ellwanger, M.~Rausch~de Traubenberg, and C.~A. Savoy,
\newblock Phys.Lett. {\bf B315}, 331 (1993), hep-ph/9307322.

\bibitem{Ellwanger:1995ru}
U.~Ellwanger, M.~Rausch~de Traubenberg, and C.~A. Savoy,
\newblock Z.Phys. {\bf C67}, 665 (1995), hep-ph/9502206.

\bibitem{Ellwanger:1996gw}
U.~Ellwanger, M.~Rausch~de Traubenberg, and C.~A. Savoy,
\newblock Nucl.Phys. {\bf B492}, 21 (1997), hep-ph/9611251.

\bibitem{Elliott:1994ht}
T.~Elliott, S.~King, and P.~White,
\newblock Phys.Lett. {\bf B351}, 213 (1995), hep-ph/9406303.

\bibitem{King:1995vk}
S.~King and P.~White,
\newblock Phys.Rev. {\bf D52}, 4183 (1995), hep-ph/9505326.

\bibitem{Franke:1995tc}
F.~Franke and H.~Fraas,
\newblock Int.J.Mod.Phys. {\bf A12}, 479 (1997), hep-ph/9512366.

\bibitem{Maniatis:2009re}
M.~Maniatis,
\newblock Int.J.Mod.Phys. {\bf A25}, 3505 (2010), 0906.0777.

\bibitem{Ellwanger:2009dp}
U.~Ellwanger, C.~Hugonie, and A.~M. Teixeira,
\newblock Phys.Rept. {\bf 496}, 1 (2010), 0910.1785.

\bibitem{Davoudiasl:2004be}
H. Davoudiasl, R. Kitano, T. Li, and H. Murayama,
\newblock {\em Phys.Lett.}, B609:117--123, 2005.

\bibitem{vanderBij:2006ne}
J.J. van~der Bij,
\newblock {\em Phys.Lett.}, B636:56--59, 2006.

\bibitem{Datta:1997fx}
A. Datta and A. Raychaudhuri,
\newblock {\em Phys.Rev.}, D57:2940--2948, 1998.

\bibitem{Schabinger:2005ei}
R. Schabinger and J.~D. Wells,
\newblock {\em Phys.Rev.}, D72:093007, 2005.

\bibitem{BahatTreidel:2006kx}
O. Bahat-Treidel, Y. Grossman, and Y. Rozen,
\newblock {\em JHEP}, 0705:022, 2007.

\bibitem{Robens:2015gla}
T. Robens and T. Stefaniak,
\newblock {\em Eur. Phys. J.}, C75:104, 2015.

\bibitem{Barger:2006sk}
V. Barger, P. Langacker, and G. Shaughnessy,
\newblock {\em Phys.Rev.}, D75:055013, 2007.

\bibitem{Barger:2007im}
V. Barger, P. Langacker, M. McCaskey, M.~J. Ramsey-Musolf, and
  G. Shaughnessy,
\newblock {\em Phys.Rev.}, D77:035005, 2008.

\bibitem{Barger:2008jx}
V. Barger, P. Langacker, M. McCaskey, M. Ramsey-Musolf, and Gabe
  Shaughnessy,
\newblock {\em Phys.Rev.}, D79:015018, 2009.

\bibitem{O'Connell:2006wi}
D. O'Connell, M.~J. Ramsey-Musolf, and Mark~B. Wise,
\newblock {\em Phys.Rev.}, D75:037701, 2007.

\bibitem{Gupta:2011gd}
R.~S. Gupta and J.~D. Wells,
\newblock {\em Phys.Lett.}, B710:154--158, 2012.

\bibitem{Ahriche:2013vqa}
A. Ahriche, A. Arhrib, and S. Nasri,
\newblock {\em JHEP}, 1402:042, 2014.

\bibitem{Coimbra:2013qq}
R.~Coimbra, M.~O.~P. Sampaio, and R.~Santos,
\newblock Eur. Phys. J. {\bf C73}, 2428 (2013), 1301.2599.

\bibitem{Chen:2014ask}
C. Chen, S.~Dawson, and I.~M. Lewis,
\newblock {\em Phys. Rev.}, D91(3):035015, 2015.

\bibitem{Profumo:2014opa}
S. Profumo, M.~J. Ramsey-Musolf, C.~L. Wainwright, and P.
  Winslow,
  \newblock {\em Phys. Rev.}, D91(3):035018, 2015.

\bibitem{Costa:2014qga}
R. Costa, A.~P. Morais, M. O.~P. Sampaio, and R. Santos,
\newblock {\em Phys. Rev.}, D92(2):025024, 2015.

\bibitem{Costa:2015llh}
  R.~Costa, M.~Mühlleitner, M.~O.~P.~Sampaio and R.~Santos,
\newblock {\em JHEP} {\bf 1606} (2016) 034.


\bibitem{JWittbrodt2016}
J.~Wittbrodt,
\newblock Master Thesis, 2016, Karlsruhe Institute of Technology  (2016).

\bibitem{Djouadi:1997yw}
A.~Djouadi, J.~Kalinowski, and M.~Spira,
\newblock Comput. Phys. Commun. {\bf 108}, 56 (1998), hep-ph/9704448.

\bibitem{Butterworth:2010ym}
J.~M. Butterworth {\em et~al.},
\newblock {THE TOOLS AND MONTE CARLO WORKING GROUP Summary Report from the Les
  Houches 2009 Workshop on TeV Colliders},
\newblock in {\em {Physics at TeV colliders. Proceedings, 6th Workshop,
  dedicated to Thomas Binoth, Les Houches, France, June 8-26, 2009}}, 2010,
  1003.1643.

\bibitem{ScannerS}
R.~Costa, R.~Guedes, M.~O.~P. Sampaio, and R.~Santos,
\newblock \textsc{ScannerS} project, 2014,
\newblock \texttt{http://scanners.hepforge.org}.

\bibitem{Horejsi:2005da}
J.~Horejsi and M.~Kladiva,
\newblock Eur. Phys. J. {\bf C46}, 81 (2006), hep-ph/0510154.

\bibitem{Klimenko:1984qx}
K.~G. Klimenko,
\newblock Theor. Math. Phys. {\bf 62}, 58 (1985),
\newblock [Teor. Mat. Fiz.62,87(1985)].

\bibitem{Coleman:1977py}
S.~R. Coleman,
\newblock Phys. Rev. {\bf D15}, 2929 (1977),
\newblock [Erratum: Phys. Rev.D16,1248(1977)].

\bibitem{Callan:1977pt}
C.~G. Callan, Jr. and S.~R. Coleman,
\newblock Phys. Rev. {\bf D16}, 1762 (1977).

\bibitem{Ferreira:2004yd}
P.~M. Ferreira, R.~Santos, and A.~Barroso,
\newblock Phys. Lett. {\bf B603}, 219 (2004), hep-ph/0406231,
\newblock [Erratum: Phys. Lett.B629,114(2005)].

\bibitem{Haber:1999zh}
H.~E. Haber and H.~E. Logan,
\newblock Phys. Rev. {\bf D62}, 015011 (2000), hep-ph/9909335.

\bibitem{Deschamps:2009rh}
O.~Deschamps {\em et~al.},
\newblock Phys. Rev. {\bf D82}, 073012 (2010), 0907.5135.

\bibitem{Mahmoudi:2009zx}
F.~Mahmoudi and O.~Stal,
\newblock Phys. Rev. {\bf D81}, 035016 (2010), 0907.1791.

\bibitem{Hermann:2012fc}
T.~Hermann, M.~Misiak, and M.~Steinhauser,
\newblock JHEP {\bf 11}, 036 (2012), 1208.2788.

\bibitem{Misiak:2015xwa}
M.~Misiak {\em et~al.},
\newblock Phys. Rev. Lett. {\bf 114}, 221801 (2015), 1503.01789.

\bibitem{Grimus:2007if}
W.~Grimus, L.~Lavoura, O.~M. Ogreid, and P.~Osland,
\newblock J. Phys. {\bf G35}, 075001 (2008), 0711.4022.

\bibitem{Grimus:2008nb}
W.~Grimus, L.~Lavoura, O.~M. Ogreid, and P.~Osland,
\newblock Nucl. Phys. {\bf B801}, 81 (2008), 0802.4353.

\bibitem{Baak:2014ora}
Gfitter Group, M.~Baak {\em et~al.},
\newblock Eur. Phys. J. {\bf C74}, 3046 (2014), 1407.3792.

\bibitem{Bechtle:2008jh}
P.~Bechtle, O.~Brein, S.~Heinemeyer, G.~Weiglein, and K.~E. Williams,
\newblock Comput. Phys. Commun. {\bf 181}, 138 (2010), 0811.4169.

\bibitem{Bechtle:2011sb}
P.~Bechtle, O.~Brein, S.~Heinemeyer, G.~Weiglein, and K.~E. Williams,
\newblock Comput. Phys. Commun. {\bf 182}, 2605 (2011), 1102.1898.

\bibitem{Bechtle:2013wla}
P.~Bechtle {\em et~al.},
\newblock Eur. Phys. J. {\bf C74}, 2693 (2014), 1311.0055.

\bibitem{Harlander:2012pb}
R.~V. Harlander, S.~Liebler, and H.~Mantler,
\newblock Comput. Phys. Commun. {\bf 184}, 1605 (2013), 1212.3249.

\bibitem{Harlander:2016hcx}
R.~V. Harlander, S.~Liebler, and H.~Mantler,
\newblock (2016), 1605.03190.

\bibitem{Khachatryan:2016vau}
ATLAS, CMS, G.~Aad {\em et~al.},
\newblock JHEP {\bf 08}, 045 (2016), 1606.02266.

\bibitem{Ferreira:2014naa}
P.~M. Ferreira, J.~F. Gunion, H.~E. Haber, and R.~Santos,
\newblock Phys. Rev. {\bf D89}, 115003 (2014), 1403.4736.

\bibitem{Ferreira:2014dya}
P.~M. Ferreira, R.~Guedes, M.~O.~P. Sampaio, and R.~Santos,
\newblock JHEP {\bf 12}, 067 (2014), 1409.6723.

\bibitem{Fontes:2014tga}
D.~Fontes, J.~C. Romao, and J.~P. Silva,
\newblock Phys. Rev. {\bf D90}, 015021 (2014), 1406.6080.

\bibitem{Krause:2016xku}
M.~Krause, M.~Muhlleitner, R.~Santos, and H.~Ziesche,
\newblock (2016), 1609.04185.

\bibitem{Barroso:2007rr}
A.~Barroso, P.~M. Ferreira, and R.~Santos,
\newblock Phys. Lett. {\bf B652}, 181 (2007), hep-ph/0702098.

\bibitem{Ivanov:2007de}
I.~P. Ivanov,
\newblock Phys. Rev. {\bf D77}, 015017 (2008), 0710.3490.

\end{thebibliography}

\end{document}